\documentclass[preprint,12pt,compress]{elsarticle}

\usepackage[colorlinks=true,linkcolor=blue,citecolor=blue,urlcolor=blue]{hyperref} 
\usepackage{amssymb}
\usepackage{multirow}
\usepackage{booktabs}
\usepackage{enumitem}
\usepackage{amsmath}
\usepackage{algorithm}
\usepackage{algpseudocode}
\usepackage{pdflscape}
\usepackage{fancyhdr} 
\usepackage{geometry}
\usepackage{graphicx}
\usepackage{subcaption}
\usepackage{verbatim}
\usepackage[normalem]{ulem}
\usepackage{wrapfig}
\usepackage{flushend}
\useunder{\uline}{\ul}{}

\journal{Applied Soft Computing}

\begin{document}

\begin{frontmatter}



\title{A Personality-Guided Preference Aggregator for Ephemeral Group Recommendation}


\author[a,b,c]{Guangze Ye}
\ead{gzye@stu.ecnu.edu.cn}
\author[c,d]{Wen Wu\corref{mycorrespondingauthor}}
\cortext[mycorrespondingauthor]{Corresponding author.}
\ead{wwu@cc.ecnu.edu.cn}
\author[e]{Liye Shi}
\ead{lyshi@ica.stc.sh.cn}
\author[f]{Wenxin Hu}
\ead{wxhu@cc.ecnu.edu.cn}
\author[d]{Xi Chen}
\ead{xchen@psy.ecnu.edu.cn}
\author[a,b,c]{Liang He}
\ead{lhe@cs.ecnu.edu.cn}

\address[a]{Lab of Artificial Intelligence for Education, East China Normal University, Shanghai, China}
\address[b]{Shanghai Institute of Artificial Intelligence for Education, East China Normal University, Shanghai, China}
\address[c]{School of Computer Science and Technology, East China Normal University, Shanghai, China} 
\address[d]{Shanghai Key Laboratory of Mental Health and Psychological Crisis Intervention, School of Psychology and Cognitive Science, East China Normal University, Shanghai, China}
\address[e]{Department of Electronic and Computer Science, Zhejiang Wanli University, Ningbo, China}
\address[f]{School of Data Science and Engineering, East China Normal University, Shanghai, China}

\begin{abstract}
Ephemeral group recommendation (EGR) aims to suggest items for a group of users who come together for the first time. Existing work typically consider individual preferences as the sole factor in aggregating group preferences. However, they neglect to take into account the importance of the individual inherent factors, such as personality, and thus fail to accurately simulate the group decision-making process. Additionally, these methods often struggle due to insufficient interactive records. To tackle these issues, a \textbf{Pe}rsonality-\textbf{G}uided Preference \textbf{A}ggregator (PEGA) is proposed, which guides the preference aggregation of group members based on their personalities, rather than relying solely on their preferences. Specifically, implicit personalities are first extracted from user reviews. Hyper-rectangles are then used to aggregate individual personalities to obtain the ``Group Personality'', which allows for the learning of personality distributions within the group. Subsequently, a personality attention mechanism is employed to aggregate group preferences, and a preference-based fine-tuning module is used to balance the weights of personality and preferences. The role of personality in this approach is twofold: (1) To estimate the importance of individual users in a group and provide explainability; (2) To alleviate the data sparsity issue encountered in ephemeral groups. Experimental results demonstrate that, on four real-world datasets, the PEGA model significantly outperforms related baseline models in terms of classification accuracy and interpretability. Moreover, empirical evidence supports the idea that personality plays a pivotal role in enhancing the performance of EGR tasks.

\end{abstract}







\end{frontmatter}



\section{Introduction}\label{section1}
Humans, as social animals, inevitably participate in various group activities, including dining, entertaining, traveling, etc. Due to the diversity of group members’ preferences, the traditional recommendation system for individuals cannot provide suggestions that satisfy all group members. As a result, the group recommendation system has emerged, which simulates the group decision-making process and finally promotes group members to reach a consensus. Groups can be categorized into persistent and ephemeral groups \citep{quintarelli2019efficiently}. Persistent groups have stable members and sufficient historical interactions, which can degenerate into a virtual individual and directly apply recommendation techniques for individuals. This paper addresses the more complex scenarios involving ephemeral groups, which are formed by users who are meeting for the first time and have minimal or no prior interactions \citep{ceh2022performance}.

Existing EGR work mainly focuses on aggregating individual users' preferences to obtain group preferences \citep{ceh2022performance}. Earlier studies mainly employ heuristics and predefined score aggregate strategies (e.g., least misery \citep{ghazarian2015enhancing}), which fail to achieve satisfactory performance due to insufficient consideration of the interactions between group members. Recently, researchers turn to data-driven approaches, including attentive models \citep{cao2019social, huang2020efficient, wang2021socially, yu2023collaborative} and graph neural network models \citep{he2022h3rec, jiang2023ktpgn, abolghasemi2024graph}. In addition, some methods improve user preference representation through regularization \citep{sankar2020groupim, Li2023SelfSupervisedGG}, while Chen et al. \cite{chen2022thinking} utilizes hypercubes to express group preferences. Although these methods have achieved satisfactory results, they still suffer from the following two limitations. 

\begin{figure}[t]
  \centering
  \begin{subfigure}[b]{0.6\textwidth}
    \includegraphics[width=\textwidth]{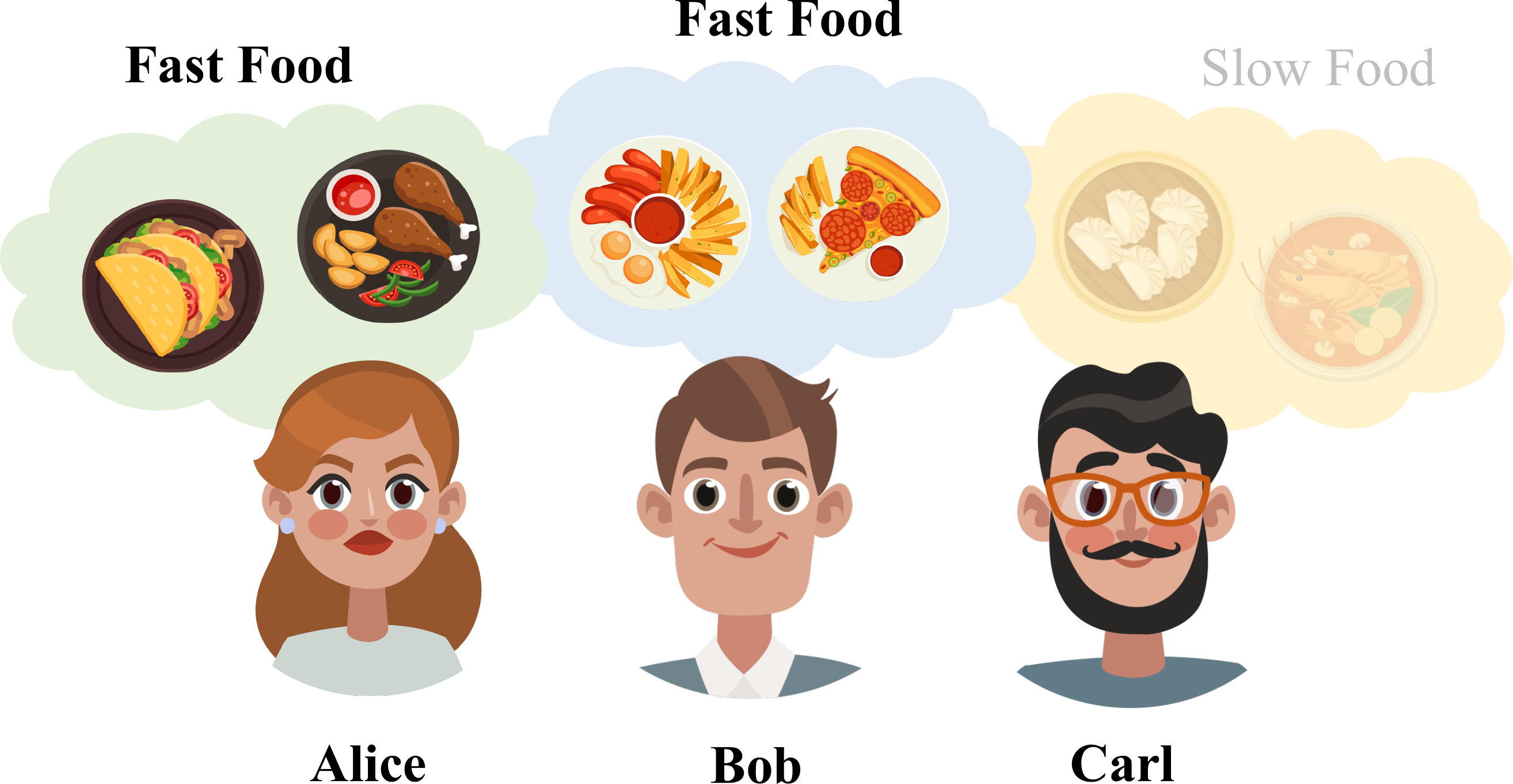}
    \caption{Without Personality}
    \label{figure1a}
  \end{subfigure}
  \quad
  \begin{subfigure}[b]{0.6\textwidth}
    \includegraphics[width=\textwidth]{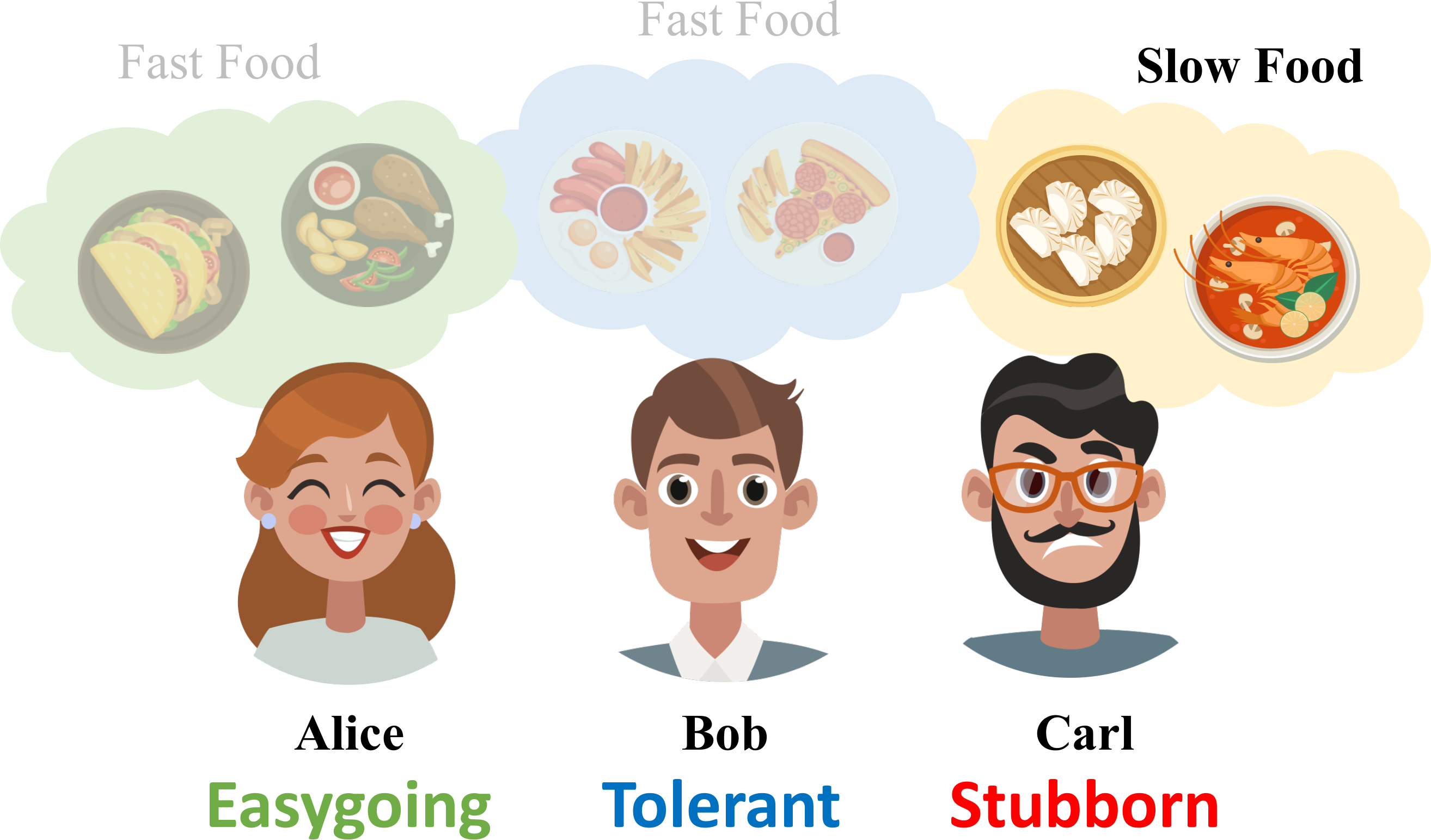}
    \caption{With Personality}
    \label{figure1b}
  \end{subfigure}
  \caption{An example that illustrates the role of personality in EGR.}
  \label{figure1}
\end{figure}

On one hand, these methods mainly use user preference information to simulate users’ importance while neglecting individual inherent factors, such as personality. Personality is described as “consistent behavior patterns and interpersonal processes originating within the individual”, which has consistently been demonstrated in the related work \citep{abolghasemi2022personality} to aid in inferring individual behavior in the group decision-making process. Meanwhile, Santos et al. \cite{santos2011personality} find that personality traits can affect group decisions by distinguishing the user roles, which makes personality a key factor in EGR. \hyperref[figure1]{Fig. 1} gives an example that illustrates the role of personality in group recommendation. Imagine that Alice, Bob, and Carl are dining out for the first time and aiming to find a restaurant. Alice and Bob are fond of fast food, while Carl prefers slow food. Following traditional preference-guided methods (\hyperref[figure1a]{Fig. 1(a)}), fast food would be the final choice since most members are fond of fast food. However, the situation changes when considering personality (\hyperref[figure1b]{Fig. 1(b)}). Carl is a stubborn person who persists in slow food, while Alice and Bob are easy-going and tolerant. Considering Carl's feelings, slow food would be the final choice. The EGR that combines personality and preference is undoubtedly more in line with the real scenario.

On the other hand, most methods suffer from the group-item interaction data sparsity issue. Existing studies \citep{cao2019social, jiang2023ktpgn} exploit external side information in the form of a social network. However, an individual's social status exhibits instability within various groups, particularly in the context of ephemeral groups. Furthermore, Self-Supervised Learning (SSL) strategies are employed to enhance user and group representations \citep{sankar2020groupim, chen2022thinking}. Nevertheless, these strategies introduce complex data structures and logical relationships, which greatly reduces the efficiency of the model \citep{Li2023SelfSupervisedGG}. Fortunately, previous studies \citep{wu2018personalizing, dhelim2020personality} have shown that incorporating personality information can enhance the performance of Collaborative Filtering (CF) in cold start scenarios. In the context of group recommendation, personality may also play a crucial role in alleviating the issue of group data sparsity. Contrary to diverse individual preferences, personality is relatively stable and will not frequently change in one’s life \citep{mairesse2007using}. Hence, with a proper amount of data, it is possible to learn multiple distribution patterns of individual personalities within a group and then guide the aggregation of individual preferences.

To address these challenges, a new solution for EGR named \textbf{Pe}rsonality-\textbf{G}uided Preference \textbf{A}ggregator (PEGA) is proposed. Firstly, the Linguistic Inquiry and Word Count (LIWC) lexicon is employed to implicitly capture users' personality traits through the linguistic features of their review texts, laying the foundation for generating personality-enhanced group representation. Secondly, the innovative concept of ``Group Personality'', derived from individual personalities, is introduced. In contrast to Chen et al. \cite{chen2022thinking}, who use the hyper-rectangle to represent group preferences, this paper choose to specifically utilize hyper-rectangle to carry stable information in the ``Group Personality''. The advantage of this choice lies in its ability to avoid the problem of hyper-rectangle becoming too large due to the diversity of member preferences, thereby preventing the distortion of information. ``Group Personality'' not only embodies the amalgamation of individual personalities within the group but also mitigates issues of data sparsity at the group level. Thirdly, a personality attention mechanism is creatively designed to replace the traditional inner and outer distance formulas in hyper-rectangles. Based on this, a personality-guided preference aggregation module is developed to aggregate the preferences of group members and generate a personality-enhanced group representation. This module not only estimates individual users’ importance in a group but also provides an explanation for group recommendation results. Finally, the inner product of the representation of the group and the candidate item is leveraged as the prediction score. Experiments conducted on four real-world datasets verify the effectiveness of the model. The experimental results show that PEGA achieves significantly higher Recall and Normalized Discounted Cumulative Gain (NDCG) scores than state-of-the-art models. 

Specifically, the main contributions of this work are as follows: 

\begin{itemize}
\item A novel ``Group Personality'' concept is introduced, which effectively addresses data sparsity issues in EGR by rapidly understanding the distribution of members' personalities from limited data, thereby accelerating the convergence of preference embeddings.

\item An innovative personality attention mechanism is developed, which enhances the explainability of EGR results by assigning appropriate weights to each group member based on their personality.

\item Implicit user personality extracted from reviews has been proven to be applicable for improving the recommendation performance of existing EGR models.

\end{itemize} 

\section{Related Work}      
\subsection{Group Recommendation}
In general, existing EGR methods often follow aggregation strategies (score aggregation and preference aggregation), which first learn members' preferences and then aggregate these preferences to infer the overall interest of the group. However, these methods often neglect other critical factors, such as personality, which play a significant role in real group decision-making. Additionally, they tend to perform suboptimally when there are few group-level interactions within ephemeral groups and lack explainability.

Concretely, score aggregation methods straightforwardly aggregated users’ scores to obtain group prediction scores by hand-craft predefined strategies. Among the strategies, average \citep{shin2009socially}, least misery \citep{ghazarian2015enhancing}, and maximum pleasure \citep{boratto2011state} were the three most popular ones. The average method takes the mean score of all members in the group, while the least misery/maximum pleasure assumes that a group is as satisfied as its least/most satisfied member. However, these methods were inflexible and ignored the real-world interactions between group members. 

Unlike score aggregation methods, preference aggregation methods adopted data-driven strategies to model group preference representation. PIT \citep{liu2012exploring} took into account the personal preferences and individual impacts of group members when determining the group preference profile. COM \citep{yuan2014generative} modeled the generative process of group activities and provided group recommendations based on that. These two probabilistic methods modeled the group generative process by considering both the personal preferences and relative influence of members, but they assumed that users have the same type of probability distribution, which is infeasible in real-world scenarios. 
To address this problem, many attention-based methods \citep{huang2020efficient, wang2021socially, yu2023collaborative, cao2018attentive} employed attention networks to explicitly model the importance of group members in the group decision-making process. For example, AGREE \citep{cao2018attentive} straightforwardly combined the attention mechanism with the neural collaborative filtering method. However, they struggled to perform well when group interaction data was severely sparse. Some research addressed the issue of data sparsity by incorporating side information \citep{cao2019social, jiang2023ktpgn, yin2019social, Deng2021KnowledgeAwareGR}, such as SIGR \citep{yin2019social} that took an attention mechanism and a bipartite graph embedding model as model blocks to learn the social-enhanced influence of users within a group. Additionally, KGAG \citep{Deng2021KnowledgeAwareGR} incorporates knowledge graphs into group recommendation tasks by recursively propagating information along the edges in the knowledge graph to learn connections among users within the group. Yet, the individual's previous social roles became invalid in the ephemeral group. 
Other methods attempted to design SSL objectives to counteract the data sparsity in group recommendation \citep{sankar2020groupim, Li2023SelfSupervisedGG, zhang2021double, chen2022thinking}. For instance, Zhang et al. \cite{zhang2021double} devised a double-scale hypergraph learning framework and employed an SSL strategy to enhance user and group representations and alleviate the data sparsity problem. GroupIM \citep{sankar2020groupim} introduced a user-group mutual information maximization scheme to jointly learn informative user and group representations, pushing the group recommendation performance to a new level. Meanwhile, SGGCF \citep{Li2023SelfSupervisedGG} explores self-supervised learning on the graph with two kinds of contrastive learning module to capture the implicit relations between groups and items. However, these SSL strategies result in suboptimal regularization outcomes because they do not consider the contextual personality effects of group members.

\subsection{Personality-based Group Recommendation}
In recommendation systems, personality can be considered a context-independent and domain-independent user profile \citep{mairesse2007using}. Many studies have applied the user’s personality traits in recommendation systems \citep{wu2018personalizing, wang2021cross}. Wu et al. \cite{wu2018personalizing} used personality to solve the cold start problem in recommendation systems and presented a personality-based greedy reranking algorithm where the personality is used to estimate the users’ diversity preferences. Ferwerda et al. \cite{wang2021cross} studied the influence of users’ personality traits on e-commerce. 

With the development of group recommendations, some earlier works discussed the importance of users’ personality traits in group decision-making. Quijano-Sanche et al. \cite{quijano2017make} argue that individuals with lower (more inclusive) personality values within a group are more likely to follow others. Zheng et al. \cite{zheng2018exploring} collected users’ Big-Five traits through Ten-Item Personality Inventory (TIPI) to divide ‘Dominators’ and ‘Followers’ in the group and used predefined score aggregation methods to aggregate ‘Dominators’ scores. Alves et al. \citep{alves2023group} collected users' Big Five personality traits through an online questionnaire and demonstrated how all five personality dimensions help predict tourist attraction choices and travel-related preferences, with only neuroticism and openness predicting travel motivations. \cite{recio2009personality} combined the Conflict Mode Weight(CMW) calculated by Thomas-Kilmann Conflict Mode Instrument (TKI) \citep{kilmann1977developing} with predefined score aggregation methods, which enhanced the influence of assertive users in the group. Additionally, Abolghasemi et al. \cite{abolghasemi2022personality} designed a consensus model based on personality traits, where these personality traits are obtained from users' responses to a TKI test consisting of 30 questions. 

However, the existing studies had two limitations: (1) They were mainly based on pre-defined heuristics strategies that failed to dynamically model the group decision-making process and lacked generalization capability; (2) It would be difficult to apply the psychological questionnaire to ephemeral groups in social media, as it would be time-consuming and impractical. Hence, this paper explores whether personality traits can be incorporated into ephemeral groups and guide the aggregation of user preferences. Instead of relying on psychological questionnaires, personality traits are implicitly captured from written review texts sourced from online social media. 

\section{Preliminaries}

To facilitate understanding, the definition of the group recommendation task is presented in this section. The sets of $M$ users, $N$ items, and $L$ groups are represented by $\mathcal{U}=\left\{u_1, u_2, \ldots, u_M\right\}$, $\mathcal{V}=\left\{v_1, v_2, \ldots, v_N\right\}$ and $\mathcal{G}=\left\{g_1, g_2, \ldots, g_L\right\}$, respectively. The $\ell$-th group $g_\ell \in \mathcal{G}$ is consisted of a set of users $\mathcal{U}_{g_\ell}=\left\{u_{\ell, 1}, u_{\ell, 2}, \ldots, u_{\ell,\left|g_\ell\right|}\right\}$, where $u_{\ell, *} \in \mathcal{U}$. The binary user-item and group-item interaction matrices are denoted by $X_U$ and $X_G$. In this work, user reviews are introduced as side information. The set of $M$ users’ review sets is denoted by $\mathcal{R}=\left\{R_{u_1}, R_{u_2}, \ldots, R_{u_M}\right\}$. Here, $R_{u_i}=\left\{r_{i, 1}, r_{i, 2}, \ldots, r_{i,\left|u_i\right|}\right\}$ is the review set of user $u_i$, where $r_{i,j}$ is $j$-th review of user $u_i$. Given a target group $g_t$, the goal is to generate a ranked list of items that $g_t$ may be interested in, which is formally defined as: 

\textbf{Input:} Users $\mathcal{U}$, items $\mathcal{V}$, groups $\mathcal{G}$, U-I interactions $X_U$, G-I interactions $X_G$ and user reviews sets $\mathcal{R}$.

\textbf{Output:} A personalized ranking function $f_{g_t}: \mathcal{V} \rightarrow \mathbb{R}$ that assigns a rating score to each item in the item set $\mathcal{V}$ for the target group $f_{g_t} \in \mathcal{G}$ \citep{sankar2020groupim, Li2023SelfSupervisedGG}. Items with higher scores are predicted to be of greater interest or relevance to the target group $f_{g_t}$.

\begin{figure}[t]
  \centering
  \includegraphics[width=1\linewidth]{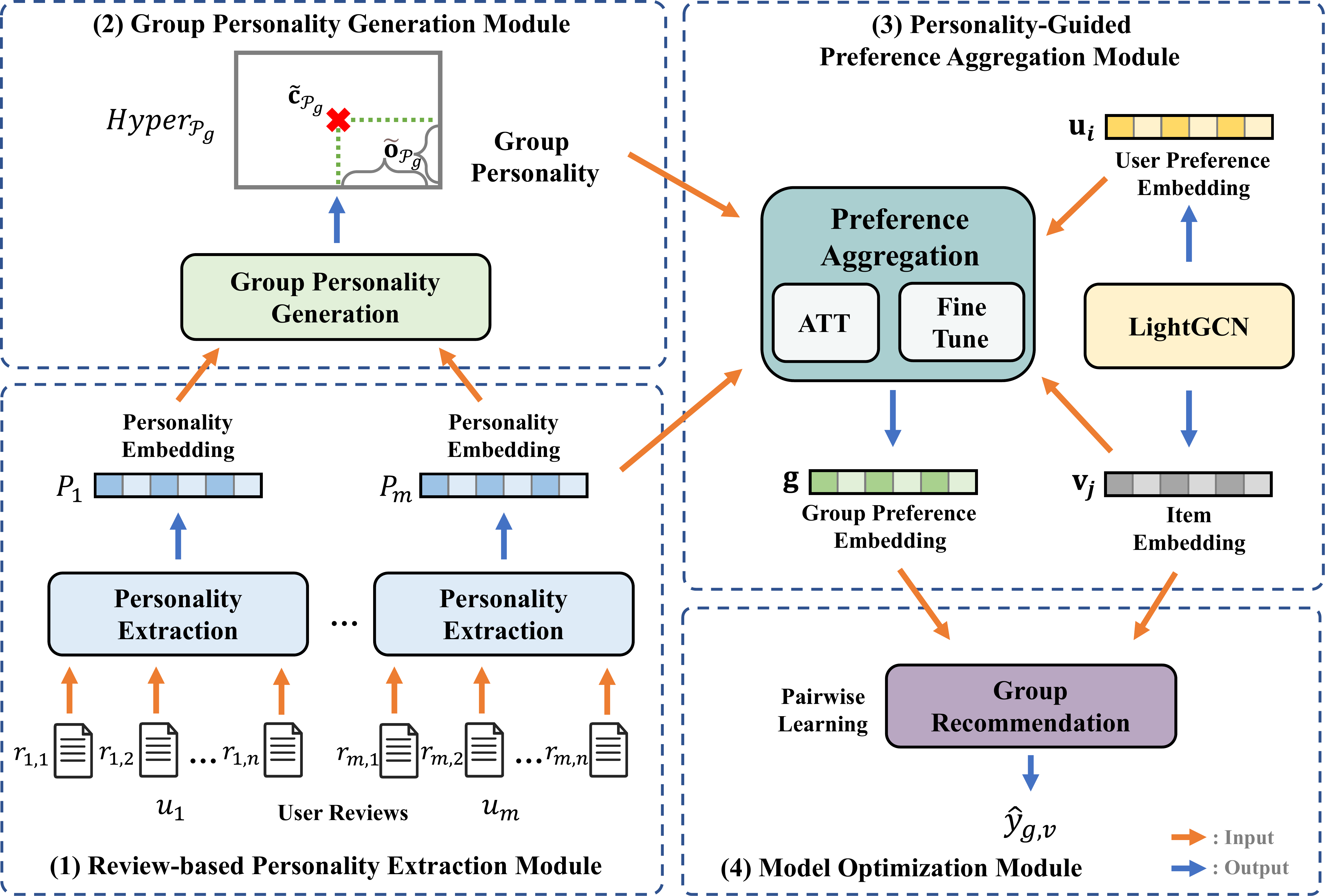}
  \caption{The framework of proposed Personality-Guided Preference Aggregator (PEGA).}
  \label{figure2}
\end{figure}

\section{Methodology}\label{section4}
In this section, the details of the PEGA model are introduced. As shown in \hyperref[figure2]{Fig. 2}, the proposed PEGA framework includes four components: (1) The \textbf{\textit{Review-based Personality Extraction Module}}, which captures users’ implicit personality traits from review texts; (2) The \textbf{\textit{Group Personality Generation Module}}, which takes individual personalities captured from (1) as input to generate ``Group Personality'' by using hyper-rectangle; (3) The \textbf{\textit{Personality-Guided Preference Aggregation Module}}, where a personality attention mechanism is first adopted to learn the user’s weight from ``Group Personality''. This weight is then fine-tuned by the user’s preference and used to aggregate group members’ preferences to obtain a personality-enhanced group representation; (4) The \textbf{\textit{Model Optimization Module}}, which applies the inner product to calculate the prediction score of the candidate item and uses a two-stage optimization scheme to optimize both user and group representations. The details of PEGA are unfolded in the following sections. 

\begin{table}[t]
\small
\centering
\begin{tabular}{@{}cll@{}}
\toprule
\multicolumn{1}{l}{Trait}          & Level                 & LIWC word Categories                           \\ \midrule
\multirow{4}{*}{O} & \multirow{2}{*}{High} & Cogproc, Insight, Cause, Tentat, Death,        \\
                   &                       & Percept, Hear, See, Anx, Space                 \\
                   & \multirow{2}{*}{Low}  & Netspeak, Family, Affect, Posemo, Reward,      \\
                   &                       & Affiliation, Focusfuture, Home, Relativ, Time  \\ \midrule
\multirow{4}{*}{C} & \multirow{2}{*}{High} & Achiev, Reward, Affiliation, Relativ, Time,           \\
                   &                       & Motion, Posemo, Work, Focusfuture, Relig        \\
                   & \multirow{2}{*}{Low}  & Negemo, Anger, Sad, Bio, Sexual,               \\
                   &                       & Body, Swear, Death, Percept, Hear              \\ \midrule
\multirow{4}{*}{E} & \multirow{2}{*}{High} & Posemo, Affiliation, Reward, Netspeak, Social, \\
                   &                       & Friend, Family, Leisure, Focusfutrue, Bio      \\
                   & \multirow{2}{*}{Low}  & Death, Work, Cogproc, Tentat, Insight,         \\
                   &                       & Differ, Cause, Risk, Negemo, Anx               \\ \midrule
\multirow{4}{*}{A} & \multirow{2}{*}{High} & Posemo, Drives, Affiliation, Reward, Achiev,           \\
                   &                       & Relativ, Time, Motion, Focusfuture, Relig      \\
                   & \multirow{2}{*}{Low}  & Negemo, Anger, Anx, Swear, Bio,               \\
                   &                       & Sexual, Body, Death, Money, Risk          \\ \midrule
\multirow{4}{*}{N} & \multirow{2}{*}{High} & Negemo, Anger, Sad, Anx, Death,                \\
                   &                       & Cogproc, Discrep, Tentat, Body, Sexual         \\
                   & \multirow{2}{*}{Low}  & Posemo, Affiliation, Reward, Achiev, Leisure,  \\
                   &                       & Relig, Netspeak, Relativ, Time, Motion         \\ \bottomrule
\end{tabular}
\caption{The strongly correlated LIWC word categories of Big-Five personality traits.}
\label{table1} 
\end{table} 

\subsection{Review-based Personality Extraction}\label{section4_1} 
In this paper, the Big-Five personality model \citep{mccrae1992introduction} is selected to represent personality traits as it is one of the most authoritative personality models. In the Big-Five framework, personality traits can be described by five dimensions: \textbf{\textit{Openness (O)}} describes a person's cognitive style and attitude towards exploring new things. \textbf{\textit{Conscientiousness (C)}} refers to controlling, managing, and regulate impulses. \textbf{\textit{Extraversion (E)}} is displayed through a higher degree of sociability, assertiveness, and talkativeness. \textbf{\textit{Agreeableness (A)}} measures the individual's attitude toward others. \textbf{\textit{Neuroticism (N)}} indicates the degree of emotional stability, impulse control, and anxiety.

The personality score of Big-Five has been proven to be related to Linguistic Inquiry and Word Count (LIWC) \citep{pennebaker1999linguistic} features \citep{kwantes2016assessing, wang2020leverage}. Therefore, the LIWC lexicon is adopted to capture users’ implicit personality traits. The LIWC lexicon contains thousands of words and phrases, which are categorized into 73 categories based on their emotional, psychological, and linguistic features. 
Similar to \citep{wang2020leverage}, the top 20 LIWC word categories that are strongly correlated with the corresponding personality traits are selected. Among these categories, 10 are associated with high-level personality traits, while the remaining 10 are associated with low-level ones. As presented in \hyperref[table1]{Table 1}, 100 lexicon categories for the 5 Big-Five dimensions are collected. For instance, a person with a high \textbf{\textit{Agreeableness (A)}} score tends to use words from the `Drives' category, such as `apology', `concern', and `dependent'. 

Then, $P=\left\{p_1, p_2, \ldots, p_{100}\right\}$ is employed as the embedding of Big-Five personality traits, where $p_i$ is the Term Frequency-Inverse Document Frequency(TF-IDF) value \citep{aizawa2003information} of lexicon words for the $i$-th lexicon category calculated over the user’s review set. Given a user $u$ and their review set $R_u$, the implicit Big-Five personality traits $P_u$ of user $u$ are calculated as:
\begin{equation}
P_u=\left\{p_{u, 1}, p_{u, 2}, \ldots, p_{u, 100}\right\}=\frac{1}{N} \sum_i^N t f_i \times \log \left(\frac{N}{d f_i}\right),
\end{equation}
where $t f_i=\left\{t f_{i, 1}, t f_{i, 2}, \ldots, t f_{i, 100}\right\}$ is the frequency of each LIWC lexicon in the $i$-th review of $R_u$, $d f_i=\left\{d f_{i, 1}, d f_{i, 2}, \ldots, d f_{i, 100}\right\}$ is the number of reviews containing words of each LIWC lexicon, and $N$ is the total number of reviews. 

\subsection{Group Personality Generation}\label{section4_2}
As previously mentioned, to overcome group interaction sparsity, the concept of ``Group Personality'' is proposed, which exhibits the personality distribution of the group. A hyper-rectangle is used to encapsulate stable information within the ``Group Personality''. Hyper-rectangle \citep{chen2022thinking} is the high-dimensional form of the rectangle where each edge of hyper-rectangle represents a real value closed interval on each personality dimension. This paradigm offers a more realistic depiction of how a group of users, each with distinct personalities, amalgamates to portray a distinctive personality distribution pattern for the group. Here, the boundary of the ``Group Personality'' can be seen as an extreme individual personality, while the geometric center exposes the primary personality tone of the entire group. Given a group $g$ and its members’ Big-Five personality traits set $\mathcal{P}_g=\left\{P_1, P_2, \ldots, P_{|g|}\right\}$, the ``Group Personality'' of $g$ is defined as follows:
\begin{equation} \label{eq2}
H_g \equiv\left\{\mathbf{v} \in \mathbb{R}^{100}: C\left(\mathcal{P}_g\right)-O\left(\mathcal{P}_g\right) \leq \mathbf{v} \leq C\left(\mathcal{P}_g\right)+O\left(\mathcal{P}_g\right)\right\},
\end{equation}
where $C\left(\mathcal{P}_g\right) \in \mathbb{R}^{100}$ is the center of the hyper-rectangle, and $O\left(\mathcal{P}_g\right) \in \mathbb{R}_{\geq 0}^{100}$ is the positive offset of the hyper-rectangle. The dimension of $C\left(\mathcal{P}_g\right)$ and $O\left(\mathcal{P}_g\right)$ are the same as with the individual personality vector $P$, which is set to 100. The center represents the essential personality of group $g$, while in the $z$-th dimension of $\mathbb{R}^{100}$, the $z$-th edge of $H_g$ spans $[C_z\left(\mathcal{P}_g\right)-O_z\left(\mathcal{P}_g\right), C_z\left(\mathcal{P}_g\right)+O_z\left(\mathcal{P}_g\right)]$, representing group $g$'s personality span in this dimension. Eq. \hyperref[eq2]{(2)} is abbreviated as a tuple for convenience, and a hyper-rectangle is computed as follows: 
\begin{equation}
H_g =\left(\mathbf{c}_g, \mathbf{o}_g\right),
\end{equation}
\begin{equation}
\mathbf{c}_g=\frac{\mathcal{P}_g^{\max }+\mathcal{P}_g^{\min }}{2},
\end{equation}
\begin{equation}
\mathbf{o}_g=\frac{\left|\mathcal{P}_g^{\max }-\mathcal{P}_g^{\min }\right|}{2},
\end{equation}
where $\mathcal{P}_g{ }^{\max }=\max \left(P_1, P_2, \ldots, P_{|g|}\right)$ and $\mathcal{P}_g{ }^{\min }=\min \left(P_1, P_2, \ldots, P_{|g|}\right)$ are element-wise boundary of ``Group Personality''. 

\subsection{Personality-Guided Preference Aggregation}
In this section, the approach to learning user-level preferences is first introduced. Next, the personality attention mechanism designed to capture the influence of individual personality within specific groups is proposed. Finally, the preference-based fine-tuning module, which fine-tunes users' weights based on their preferences, is described. 

\subsubsection{User-Level Preferences Learning}
This paper focus on aggregating individual preferences under the guide of personality traits. LightGCN \citep{he2020lightgcn} is emploied to learn user and item embedding, using the message-passing mechanism to facilitate modeling higher-order collaborative signals. The embeddings are as follows:
\begin{equation}
U, V=\operatorname{LightGCN}\left(X_U\right),
\end{equation}
where $X_U$ is user-item interaction metric, $\operatorname{LightGCN}$ is the functional training process of the model LightGCN, $U=\left\{\mathbf{u}_1, \mathbf{u}_2, \ldots, \mathbf{u}_M\right\}$, $\mathbf{u}_i \in \mathbb{R}^{M \times d}$ and $V=\left\{\mathbf{v}_1, \mathbf{v}_2, \ldots, \mathbf{v}_N\right\}$, $\mathbf{v}_i \in \mathbb{R}^{N \times d}$ are $d$-dimensional embedding sets of user and item, respectively. 

\subsubsection{Personality Attention Mechanism}
As described in \hyperref[section4_2]{Section 4.2}, the geometric center represents the primary personality tone of the group, whereas the boundary of ``Group Personality'' is considered to represent an extreme individual personality. Differing from Chen et al. \cite{chen2022thinking}, simple calculation of the inner and outer distances of hyper-rectangle cannot measure the importance of individual personality $P_t$ in relation to ``Group Personality'' $H_g$. This is because it cannot assert that individual personalities closer to the tone of ``Group Personality'' are more important. In fact, it is necessary to simultaneously measure the importance of individual personality in relation to both the group's main personality tone and its extreme personalities, rather than distance. In light of this, an approach is proposed that measures the individual personality importance $\tilde{\alpha}(g, t)$ via: 
\begin{equation}
\label{eq7}
\tilde{\alpha}(g, t)=(1-\gamma)f_{O}(H_g, P_t) + \gamma f_{C}(H_g, P_t)
\end{equation} 
where the functions $f_{O}$ and $f_{C}$ measures the importance of the individual personality $P_t$ in relation to the group's extreme personality and its main personality tone, respectively, which is balanced by a fixed coefficient $\gamma$. Taking the group $g$'s personality $H_g$ and individual personality $P_t$ of user $u_t \in \mathcal{U}_g$ as inputs, $f_{O}$ is formulated as follows:
\begin{equation} 
f_{O}(H_g, P_t)=\operatorname{Softmax}(\alpha_O(g, t))=\frac{\exp \alpha_O(g, t)}{\sum_{t^{\prime} \in \mathcal{U}_g} \exp \alpha_O\left(g, t^{\prime}\right)}, 
\end{equation} 
\begin{equation}
\alpha_O(g, t)=\mathbf{H}^T \operatorname{Tanh}\left(\mathbf{W}_q \mathbf{o}_g+\mathbf{W}_k (P_t-\mathbf{c}_g)+\mathbf{b}\right),
\end{equation}  
where the difference between the individual personality $P_t$ and the center $\mathbf{c}_g$ of the ``Group Personality'' $H_g$ is employed as the query vector, while the offset $\mathbf{o}_g$ of the ``Group Personality'' $H_g$ is employed as the key vector. Similarly, $f_{C}$ is formulated as follows:
\begin{equation} 
f_{C}(H_g, P_t)=\operatorname{Softmax}(\alpha_C(g, t))=\frac{\exp \alpha_C(g, t)}{\sum_{t^{\prime} \in \mathcal{U}_g} \exp \alpha_C\left(g, t^{\prime}\right)}, 
\end{equation} 
\begin{equation}
\alpha_C(g, t)=\mathbf{H}^T \operatorname{Tanh}\left(\mathbf{W}_q \mathbf{c}_g+\mathbf{W}_k P_t+\mathbf{b}\right),
\label{eq11}
\end{equation} 
where the center $\mathbf{c}_g$ of ``Group Personality'' $H_g$ is employed as the query vector, while the individual personality $P_t$ is employed as the key vector. 

In the above formulas, $\mathbf{W}_q \in \mathbb{R}^{100 \times 100}$ and $\mathbf{W}_k \in \mathbb{R}^{100 \times 100}$ are trainable weight matrices that convert ``Group Personality'' and user personality to the hidden layer, respectively. Additionally, $\mathbf{b} \in \mathbb{R}^{100}$ is the bias vector. Tanh serves as the activation function for the personality attention network, and $\mathbf{H} \in \mathbb{R}^{100}$ is the weight vector that projects the hidden layer to the scores $\alpha_O(g, t)$ and $\alpha_C(g, t)$. Finally, the Softmax function is adopted to obtain normalized attention weights. 

\subsubsection{Preference-based Fine-tuning}
Considering that in some cases, personal preferences may enhance an individual's influence, PEGA further considers the personal preferences of candidate items. For example, a vegetarian may exert great effort in persuading others to embrace vegetarianism, while choosing to remain silent on other topics or in different situations. The preference-based fine-tuning module is introduced to balance relatively stable personality and variable preference. Specifically, the similarity calculated between the user and the candidate item is used to properly enhance the user’s influence. Firstly, the personality traits of user $u_t$ are incorporated into their preference embedding to obtain a more comprehensive embedding
$\mathbf{u}_t^{\prime}$ of user $u_t$ as follows:
\begin{equation}
\mathbf{u}_t^{\prime}=\mathbf{u}_t \parallel P_t,
\end{equation}
where $\parallel$ is the concatenation operation, $\mathbf{u}_t$ is the embedding of user $u_t$, and $P_t$ is the personality traits of $u_t$. For the $i$-th candidate item $v_i$, the user $u_t$'s preference can be calculated as follows:
\begin{equation}
\beta(i, t)=\mathbf{v}_i{ }^T \mathbf{w} \mathbf{u}_t{ }^{\prime},
\end{equation}
\begin{equation}
\tilde{\beta}(i, t)=\operatorname{Softmax}(\beta(i, t))=\frac{\exp \beta(i, t)}{\sum_{t^{\prime} \in u_g} \exp \beta\left(i, t^{\prime}\right)},
\end{equation}
where $\mathbf{v}_i \in \mathbb{R}^d$ is the embedding of candidate item $v_i$, and $\mathbf{w} \in \mathbb{R}^{d \times\left|\mathbf{u}_t{ }^{\prime}\right|}$ is a trainable matrix that projects $\mathbf{u}_t^{\prime}$ into the same dimensional space as $\mathbf{v}_i$.
Finally, a weighted sum is performed on the embeddings of group $g$’s member users $U$, and the group preference $\mathbf{g}$ is abstracted as follows:
\begin{equation}
\mathbf{g}=\sum_{t \in u_g} \delta(g, t, i) \mathbf{u}_t,
\end{equation}
\begin{equation}
\delta(g, t, i)=(1 - \lambda) \tilde{\alpha}(g, t)+\lambda \tilde{\beta}(i, t),
\label{eq16}
\end{equation}
where $\delta(g, t, i)$ is the overall influence weight of user $u_t$ in group $g$ towards candidate item $v_i$, and $\lambda$ is a fixed coefficient that balances the importance of an individual’s personality and preference.

\subsection{Model Optimization}
To optimize the parameters of PEGA, a joint training approach is adopted using Bayesian Personalized Ranking (BPR) \citep{rendle2009bpr} pairwise learning. BPR pairwise learning aims to maximize the score difference between positive and negative items. Specifically, user and item embeddings are optimized by minimizing the user-level BPR pairwise loss $\mathcal{L}_{u s e r}$ as follows: 
\begin{equation}
\mathcal{L}_{u s e r}=-\sum_{\left(u, v_p, v_n\right) \in \mathcal{O}} \log \sigma\left(\hat{y}_{u, v_p}-\hat{y}_{u, v_n}\right),
\end{equation}
where $\mathcal{O}$ is the set of user training instances. Each instance $\left(u, v_p, v_n\right)$ contains a positive item $v_p$ that the user $u$ has interacted with and a negative item $v_n$ that the user $u$ 
hasn’t interacted with yet. $\sigma$ is the sigmoid function, and $\hat{y}_{u, v_p}$ and $\hat{y}_{u, v_n}$ are the predicted score for $v_n$ and $v_p$ which are calculated from:
\begin{equation}
\hat{y}_{u, v}=\mathbf{u} \cdot \mathbf{v},
\end{equation}
where element-wise dot product $\cdot$ is applied to the user $u$'s embedding $\mathbf{u}$ and the item $v$'s 
embedding $\mathbf{v}$. Similarly, the group preference representation is optimized by minimizing the combined loss $ \mathcal{L} $ as follows:

\begin{equation}
\mathcal{L} = \mathcal{L}_{g r o u p} + \mu \mathcal{L}_{u s e r} (\mu \geq 0)
\label{eq19}
\end{equation}
\begin{equation}
\mathcal{L}_{g r o u p}=-\sum_{\left(g, v_p, v_n\right) \in \mathcal{O}^{\prime}} \log \sigma\left(\hat{y}_{g, v_p}-\hat{y}_{g, v_n}\right),
\end{equation}
where $\mathcal{L}_{g r o u p}$ is the group-level BPR pairwise loss, and hyper-parameter $\mu$ is the weight of the user ranking loss. $\mathcal{O}^{\prime}$ is the set of group training instances, where each instance $\left(g, v_p, v_n\right)$ contains a positive item $v_p$ that the group $g$ has interacted with and a negative item $v_n$ that group $g$ hasn’t interact with yet. $\sigma$ is the sigmoid function, $\hat{y}_{g, v_p}$ and $\hat{y}_{g, v_n}$ are the predicted score for $v_n$ and $v_p$ which are calculated from:
\begin{equation}
\hat{y}_{g, v}=\mathbf{g} \cdot \mathbf{v},
\end{equation}
where element-wise dot product $\cdot$ is applied to the group $g$'s embedding $\mathbf{g}$ and the item $v$'s 
embedding $\mathbf{v}$. Model parameters are optimized using the Adam optimizer. The entire training process is repeated until the value of $\mathcal{L}$ is sufficiently small.

\section{Experiments}
In this section, the aim is to answer the following research questions (RQs):
\begin{itemize}
\item \textbf{RQ1}: Does the proposed PEGA approach outperform state-of-the-art models?
\item \textbf{RQ2}: What are the benefits of PEGA's major components?
\item \textbf{RQ3}: How does personality information extracted from the review data affect the performance of the baseline models? 
\item \textbf{RQ4}: What is the impact of the hyper-parameters of PEGA?
\item \textbf{RQ5}: How does PEGA perform with different group sizes?
\item \textbf{RQ6}: Can individual personality in PEGA help explain the results of group recommendation? 
\item \textbf{RQ7}: Can ``Group Personality'' in PEGA help alleviate the data sparsity issue of group recommendation?
\end{itemize}

\begin{figure}[t]
  \centering
  \includegraphics[width=1\linewidth]{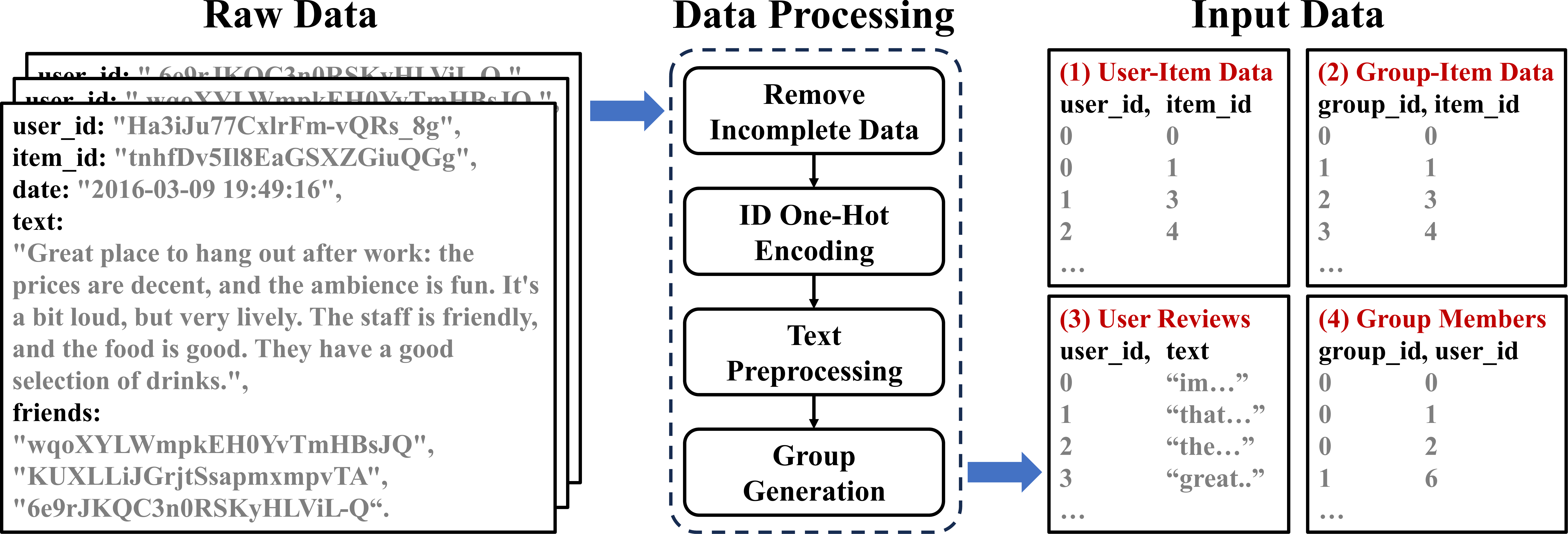}
  \caption{Data samples and data processing workflow in EGR.}
  \label{figure3}
\end{figure}

\subsection{Datasets}\label{section5_1}
In EGR, datasets usually contain user IDs, group IDs, and item IDs as features. Since PEGA extracts personality traits from user reviews, the dataset used in this paper also includes review text features. Specifically, four new datasets (i.e., Amazon-Simi, Amazon-Rand, Yelp2017-2018, and Yelp2019-2020) are built based on the original \textbf{Yelp}$\footnote{https://www.yelp.com/dataset}$ and \textbf{Amazon}$\footnote{https://jmcauley.ucsd.edu/data/amazon/}$ datasets, 
employing widely adopted group data construction approaches \citep{sankar2020groupim, Deng2021KnowledgeAwareGR, zhou2023multi}. For clarity, \hyperref[figure3]{Fig. 3} illustrates data samples and the data processing flow in EGR. The left shows raw data samples, the middle illustrates the data processing flow, and the right presents the final input data for the model. Here, \textbf{`user\_id'} and \textbf{`item\_id'} are independent and meaningless strings that cannot be directly used. After one-hot encoding, they form the user-item interaction table. The cleaned review \textbf{`text'} forms the user-text table, while \textbf{`date'} and \textbf{`friends'} are used only for constructing group interactions and do not participate in model training. The subsequent paragraphs provide further details on dataset construction.

Specifically, the first two datasets, Amazon-Simi and Amazon-Rand, are extracted from Amazon Books, a subset of Amazon where the merchandise category is books. Amazon Books contains 22 million explicit ratings (ranging from 1 to 5) and 8 million reviews, spanning from May 1996 to July 2014. Following the approaches in \citep{Deng2021KnowledgeAwareGR, zhou2023multi}, the groups are divided into two types: \textit{similar} and \textit{random}, based on how similar group members are in their ratings, and the corresponding Amazon-Simi and Amazon-Rand datasets are generated. For Amazon-Simi, the group members share similar interests, and Pearson Correlation (PCC) \citep{pearson1896vii} is used to calculate the similarity between them. Specifically, the PCC threshold value is set at 0.27, in accordance with \citep{Deng2021KnowledgeAwareGR}. This threshold implies that the PCC between each pair of users in the same group must be higher than 0.27. In contrast, for Amazon-Rand, the groups are randomly structured without any constraints on user-to-user similarity among its members. Following prior work \citep{Deng2021KnowledgeAwareGR, yin2019social, zhou2023multi}, the ground-truth item is defined as one where every member in the group gives a rating higher than 4. 

The Yelp dataset presents a real-world collection of user check-in and comment data pertaining to local businesses. With a vast compilation of 7 million reviews from 2 million users across 150,000 businesses, the dataset spans from February 2005 to January 2022. To ensure manageable data volumes and minimize potential experimental biases, interaction data with missing timestamps are removed, and data from January 2017 to December 2018 and from January 2019 to December 2020 are specifically extracted. This results in two distinct datasets, named Yelp2017-2018 and Yelp2019-2020, respectively. 
In constructing group interactions within both datasets, the method proposed by Yin et al. \cite{yin2019social} and subsequently widely used \citep{sankar2020groupim, Deng2021KnowledgeAwareGR, guo2021hierarchical, chen2022thinking, Li2023SelfSupervisedGG} is employed. 
This approach involves combining information from check-ins and social network connections. More specifically, a group interaction is formed when a set of users who are friends on the social network collectively check-in at the same business within a 15-minute time frame. 

\begin{table}[t]
\small
\centering
\begin{tabular}{lrrrr}
\hline
Dataset                    & \begin{tabular}[c]{@{}r@{}}Amazon-\\ Simi\end{tabular} & \begin{tabular}[c]{@{}r@{}}Amazon-\\ Rand\end{tabular} & \begin{tabular}[c]{@{}r@{}}Yelp2017-\\ 2018\end{tabular} & \begin{tabular}[c]{@{}r@{}}Yelp2019-\\ 2020\end{tabular}  \\ \hline
\# Users                   & 33,589                                                 & 44,843                                                 & 19,007   & 13,543    \\
\# Items                   & 24,806                                                 & 26,588                                                 & 38,665   & 18,699    \\
\# Groups                  & 13,145                                                 & 13,332                                                 & 33,782   & 32,903    \\
\# U-I interactions   & 191,366                                                & 207,777                                                & 330,956  & 239,178   \\
\# G-I interactions & 15,053                                                 & 15,282                                                & 34,830   & 19,441    \\
Avg. \# items per user     & 5.70                                                   & 4.63                                                   & 17.41    & 17.66    \\
Avg. \# items per group    & 1.15                                                   & 1.15                                                   & 1.03     & 1.04     \\
Avg. group size            & 5.55                                                   & 6.03                                                   & 4.47     & 4.22     \\ \hline
\end{tabular}
\caption{Dataset Statistics.}
\label{table2}
\end{table} 

\begin{figure}[htbp]
    \centering
    \subfloat[Amazon]{%
        \includegraphics[width=0.48\textwidth]{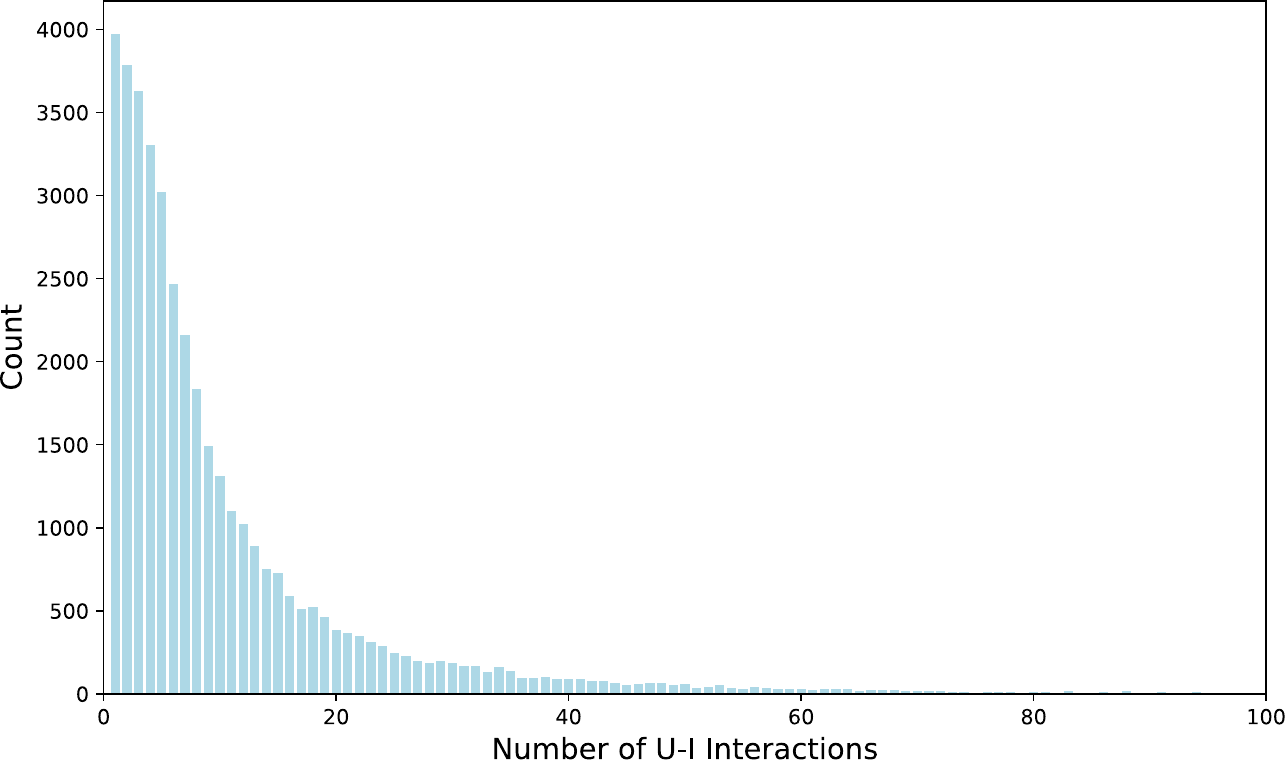}
    }
    \subfloat[Yelp]{%
        \includegraphics[width=0.48\textwidth]{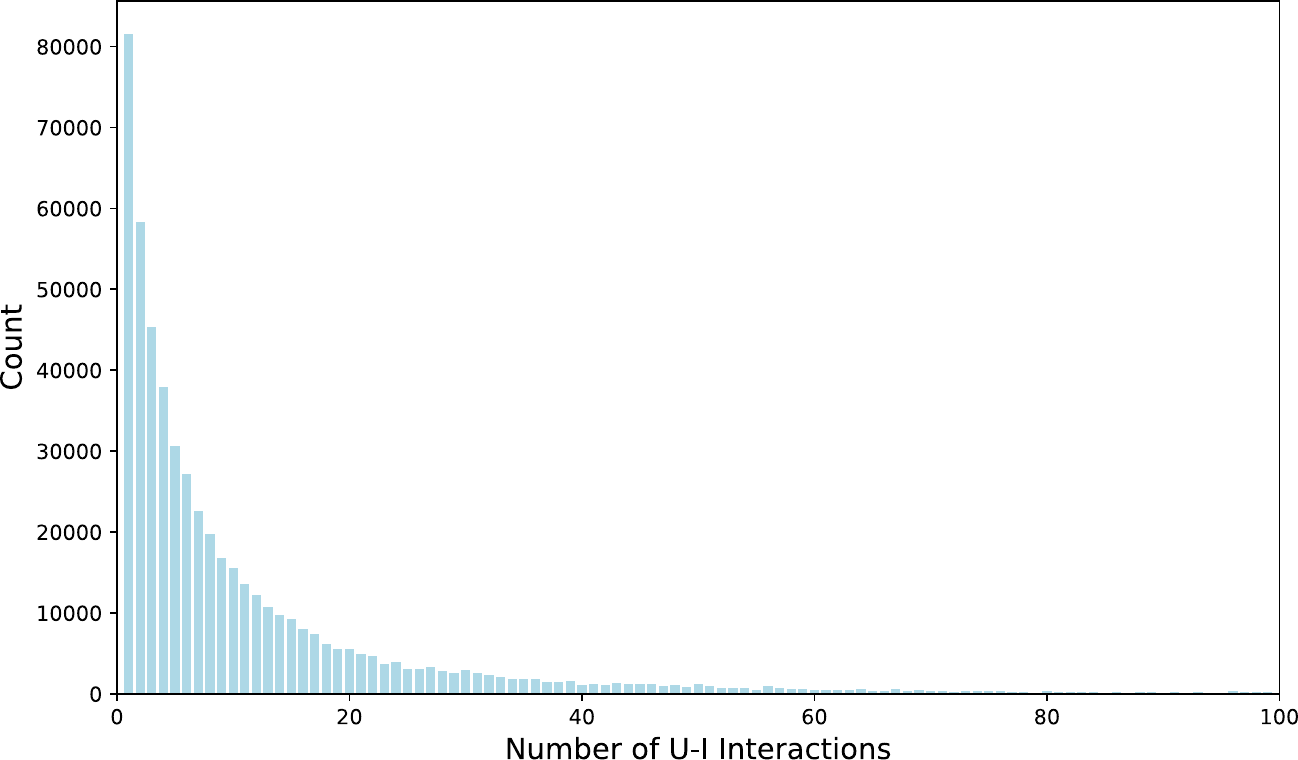}
    }
    \vskip\baselineskip
    \subfloat[Amazon-Simi]{%
        \includegraphics[width=0.48\textwidth]{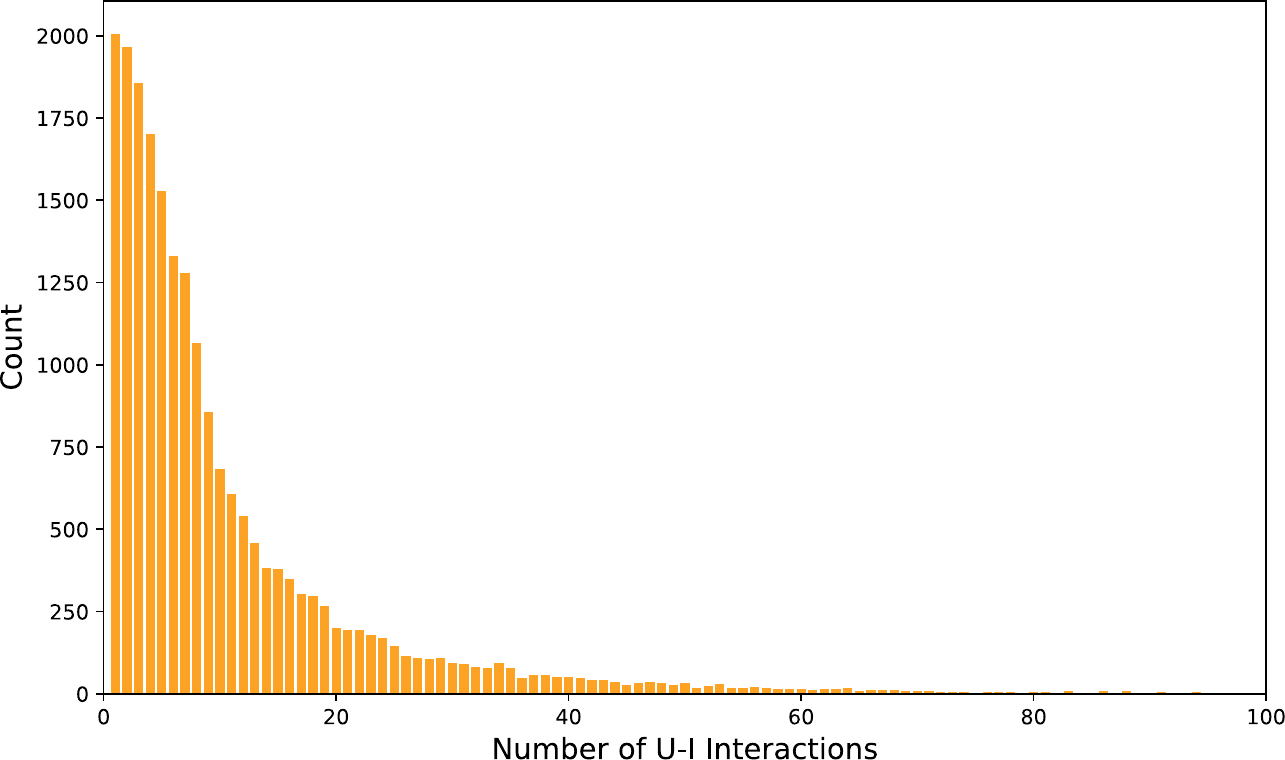}
    }
    \subfloat[Yelp2017-2018]{%
        \includegraphics[width=0.48\textwidth]{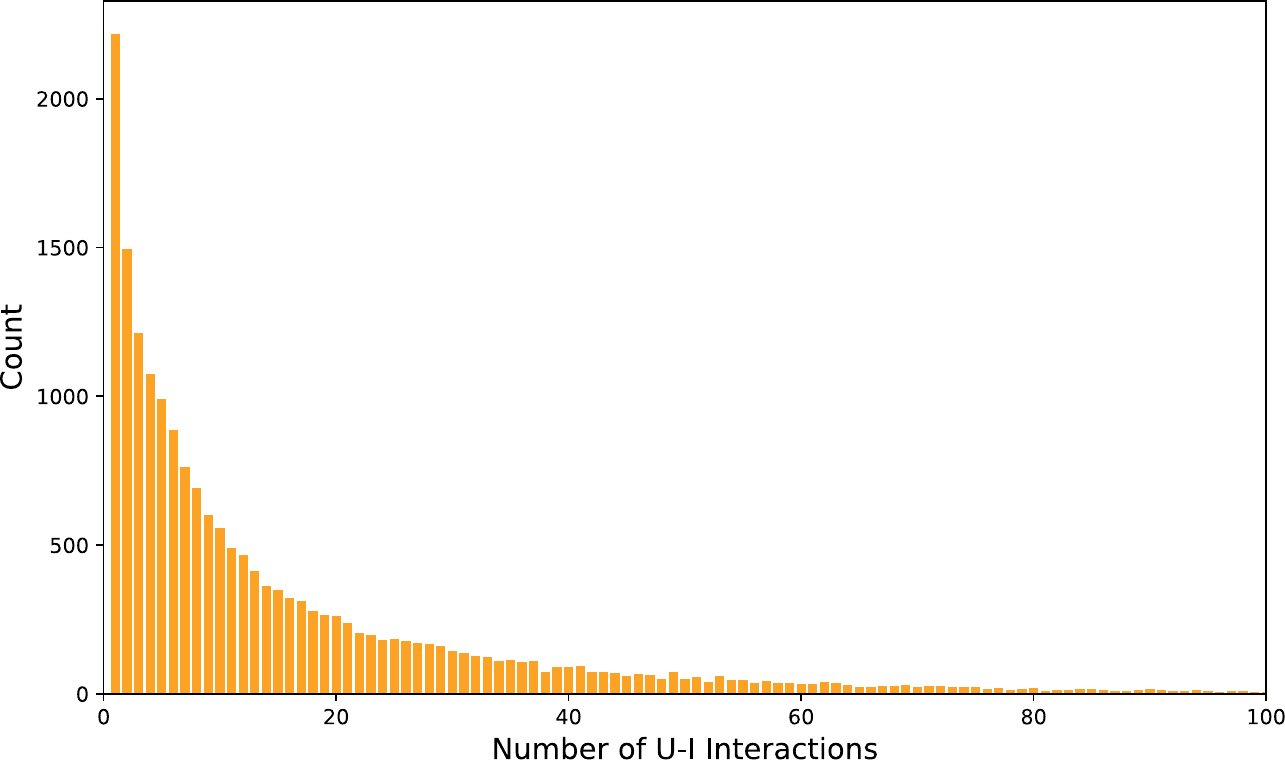}
    }
    \vskip\baselineskip
    \subfloat[Amazon-Rand]{%
        \includegraphics[width=0.48\textwidth]{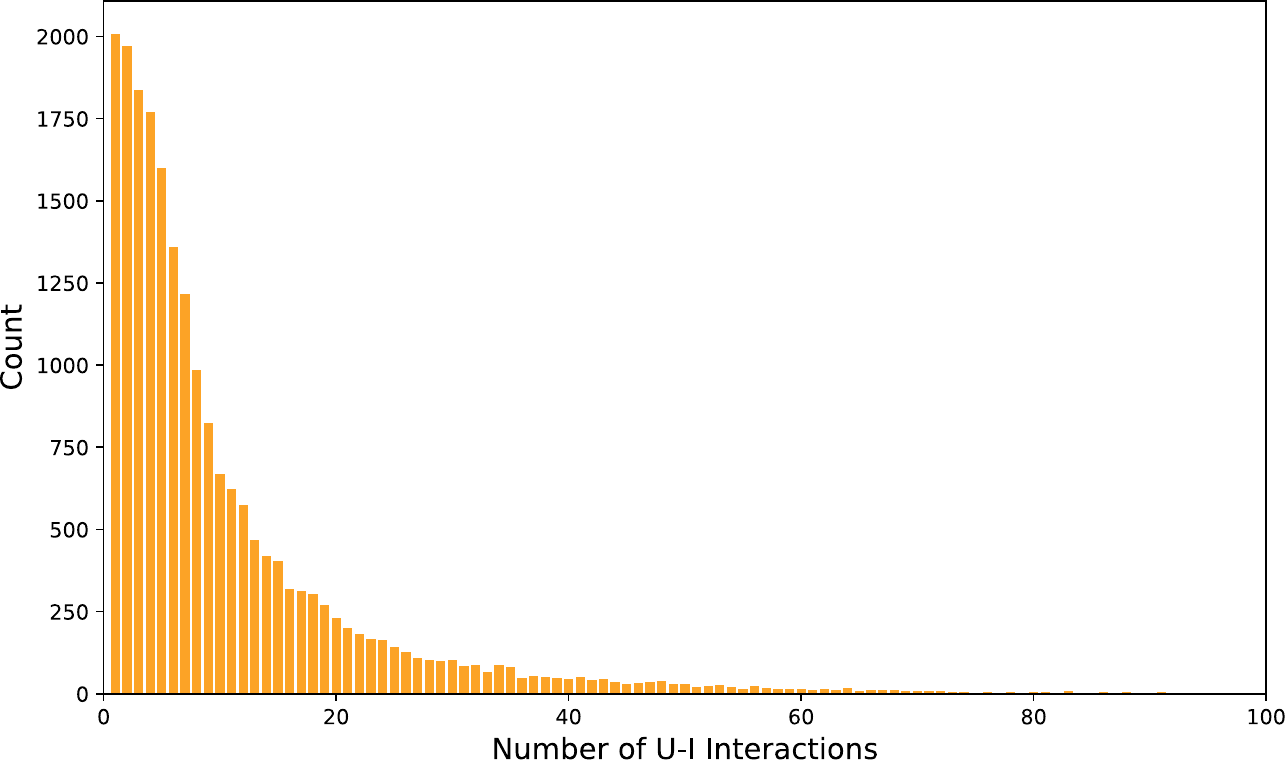}
    }
    \subfloat[Yelp2019-2020]{%
        \includegraphics[width=0.48\textwidth]{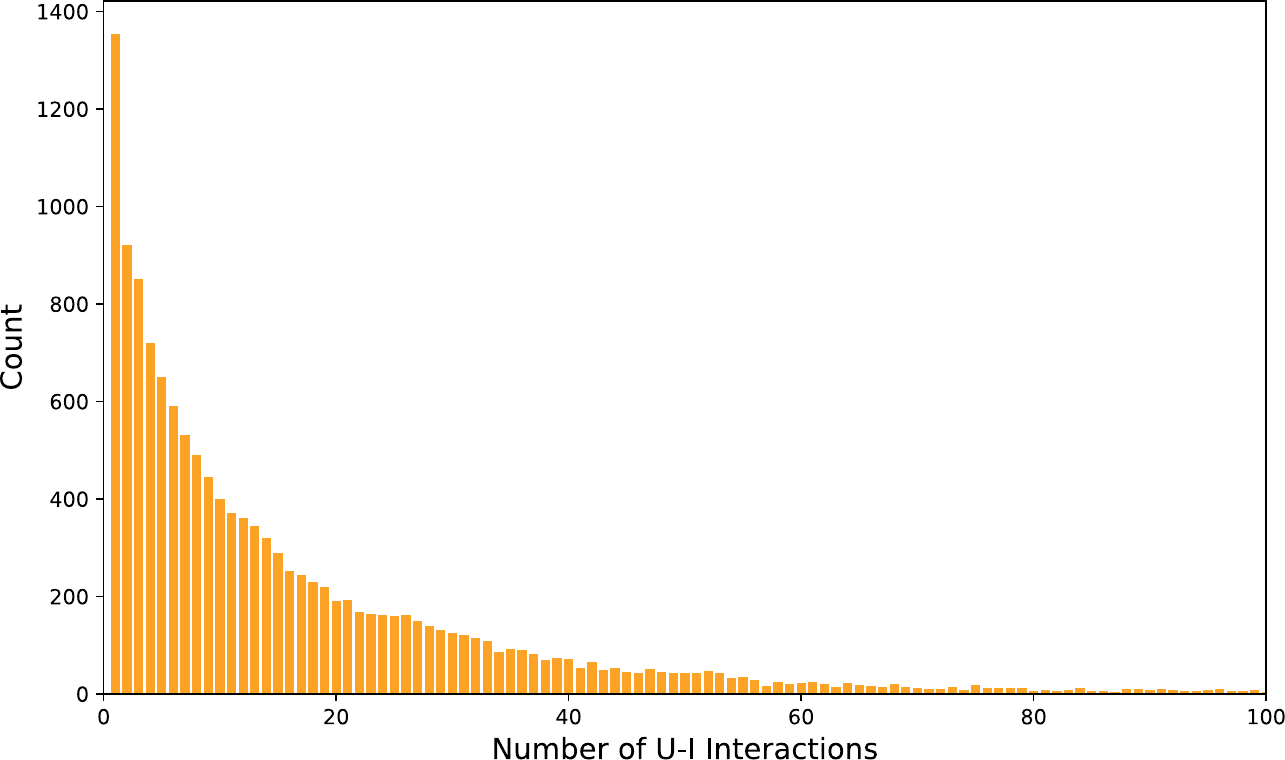}
    }
    \caption{Distribution of U-I interaction data in the four datasets compared to the original dataset.}
    \label{figure4}
\end{figure}

\hyperref[table2]{Table 2} shows the key statistics of four datasets, in which the groups interact with only one item on average because they are ephemeral, as mentioned before. Additionally, to further validate these four datasets, the distribution of user-item interactions is analyzed and visualized. \hyperref[figure4]{Fig. 4} illustrates this distribution, which is crucial for recommender system datasets. The first row of images displays the user-item interaction distributions for the raw Amazon and Yelp data, while the second and third rows show the distributions for the four newly constructed datasets. The figures reveal that the distributions of the new datasets closely align with those of their raw datasets, thus confirming the effectiveness of the datasets.

\subsection{Baselines}
The proposed PEGA model is compared with the following state-of-the-art models, which can be broadly divided into \textit{predefined score aggregation methods} (i.e., NFC-AVG, NCF-LM and NCF-MAX), \textit{data-driven preference-guided aggregation methods} (i.e., AGREE, CubeRec, SGGCF and GroupIM), and \textit{rule-based personality-guided aggregation method} (i.e., NFC-Big5). 

\begin{itemize}
\item \textbf{NCF-AVG}, \textbf{NCF-LM}, and \textbf{NCF-MAX} are based on the state-of-the-art individual recommendation model NCF \citep{he2017neural}, which replaces the inner product with a neural architecture to complete the recommendation task. NCF is combined with three predefined score aggregation strategies, including average (AVG), least misery (LM), and maximum satisfaction (MAX).  
\item \textbf{AGREE} \citep{cao2018attentive} is an attention-based neural group recommender that calculates item-specific attention weight of use by jointly training user-item and group-item interactions.
\item \textbf{CubeRec} \citep{chen2022thinking} is an ephemeral group recommender that uses hypercube vector space to replace point embeddings of user's preference to represent group preference.
\item \textbf{S$^2$-HHGR} \citep{zhang2021double} is an ephemeral group recommender using a hybrid approach that combines hierarchical hypergraph learning with self-supervised learning to enhance group representation.
\item \textbf{SGGCF} \citep{Li2023SelfSupervisedGG} is an ephemeral group recommender that leverages self-supervised learning with two kinds of contrastive learning modules and a unified user-centered graph.
\item \textbf{GoupIM} \citep{sankar2020groupim} is a recommender architecture-agnostic framework that maximizes user-group mutual information when regularizing the representation space of users and groups.
\item \textbf{NCF-Big5} \citep{Rossi2016SocialUA} is a rule-based group recommender that takes into account users’ agreeableness factor of the Big-Five personality traits. It utilizes the NCF model to derive user and item embeddings and assigns higher weights to users with higher agreeableness scores when aggregating the group preference score. The first 20 dimensions of $P_u$ (which include the user's implicit agreeableness dimension information) are mapped to a discrete value between 1 and 5 to obtain the personal value mentioned in \cite{Rossi2016SocialUA}.

\end{itemize}

\subsection{Experimental Setup}
\subsubsection{Evaluation Metrics}
Following \citep{sankar2020groupim}, two widely adopted evaluation metrics are used to evaluate the performance of PEGA: recall at rank K(R@K) and NDCG at rank K(N@K) where K=$\left\{20, 50\right\}$. In addition, Value Improvement Percentage (VIP) is used to measure the value improvement percentage of PEGA compared to other methods. 
\begin{equation}
V I P=\frac{{V a l u e}_{{PEGA}}-{V a l u e}_{{c o m p a r e d m e t h o d}}}{{V a l u e}_{{c o m p a r e d m e t h o d}}}
\end{equation}
The datasets are split into training, validation, and testing sets according to the proportion of 8:1:1. Five-fold cross-validation is performed for each dataset to avoid any biases. Since the data are not normally distributed, the permutation test \citep{nichols2002nonparametric} is adopted for significance tests. 
\subsubsection{Implementations}
The models mentioned above are reproduced, and the grid search method is used to obtain optimal hyperparameters. Specifically, the learning rate and dropout rate are searched in $\left\{0.01, 0.001, 1e-4\right\}$ and $\left\{0, 0.3, 0.5, 0.7\right\}$, respectively. In PEGA, the latent dimension $d$, the balance coefficient of personality attention weights $\gamma$, the weight of the user ranking loss $\mu$, and the balance coefficient of personality and preference $\lambda$ are fixed as 256, 0.5, 0.5, and 0.3, respectively. The Adam optimizer is adopted for all models, and 5 negative items are sampled for each ground truth item. All experiments are conducted on a single Nvidia GeForce RTX 4090 GPU with PyTorch implementations on the Linux platform.

\fancypagestyle{mylandscape}{
  \fancyhf{}              
  \renewcommand{\headrulewidth}{0pt}   
  \renewcommand{\footrulewidth}{0pt}   
  \fancyfoot[R]{\thepage}
}

\newpage
\newgeometry{margin=1in}
\begin{landscape}
\thispagestyle{mylandscape}

\begin{table}[]
\setlength{\tabcolsep}{0.3mm}
\small
\centering
\begin{tabular}{@{}lllll|cccc|cccc|cccc@{}}
\hline
\textbf{Dataset}     & \multicolumn{4}{c|}{\textbf{Amazon-Simi}}                                                  & \multicolumn{4}{c|}{\textbf{Amazon-Rand}}                                                  & \multicolumn{4}{c|}{\textbf{Yelp2017-2018}}                                                     & \multicolumn{4}{c}{\textbf{Yelp2019-2020}}                                                                                                         \\
\textbf{Metric}      & \textbf{N@20}   & \textbf{N@50}   & \textbf{R@20}   & \multicolumn{1}{l|}{\textbf{R@50}}   & \textbf{N@20}   & \textbf{N@50}   & \textbf{R@20}   & \multicolumn{1}{l|}{\textbf{R@50}}   & \textbf{N@20}   & \textbf{N@50}   & \textbf{R@20}   & \multicolumn{1}{l|}{\textbf{R@50}}   & \multicolumn{1}{c}{\textbf{N@20}} & \multicolumn{1}{c}{\textbf{N@50}} & \multicolumn{1}{c}{\textbf{R@20}} & \multicolumn{1}{c}{\textbf{R@50}} \\ \hline
\multicolumn{17}{c}{\textbf{Predefined Score Aggregators}}                                                                                                                                                                                                                                                                                                                                                                                                   \\ \hline
\textbf{NCF+AVG}     & 0.234           & 0.248           & 0.376           & \multicolumn{1}{l|}{0.446}           & 0.169           & 0.184           & 0.305           & \multicolumn{1}{l|}{0.384}           & 0.164           & 0.177           & 0.268           & \multicolumn{1}{l|}{0.330}           & 0.214                             & 0.227                             & 0.340                             & 0.416                             \\
\textbf{NCF+LM}      & 0.278           & 0.302           & 0.492           & \multicolumn{1}{l|}{0.613}           & 0.201           & 0.227           & 0.404           & \multicolumn{1}{l|}{0.539}           & 0.096           & 0.106           & 0.158           & \multicolumn{1}{l|}{0.208}           & 0.135                             & 0.145                             & 0.211                             & 0.265                             \\
\textbf{NCF+MAX}     & 0.259           & 0.287           & 0.397           & \multicolumn{1}{l|}{0.521}           & 0.189           & 0.206           & 0.326           & \multicolumn{1}{l|}{0.443}           & 0.123           & 0.137           & 0.234           & \multicolumn{1}{l|}{0.305}           & 0.147                             & 0.164                             & 0.298                             & 0.386                             \\ \hline
\multicolumn{17}{c}{\textbf{Preference-Guided Aggregators}}                                                                                                                                                                                                                                                                                                                                                                                                 \\ \hline
\textbf{AGREE}       & 0.289           & 0.502           & 0.613           & \multicolumn{1}{l|}{0.692}           & 0.383           & 0.530           & 0.669           & \multicolumn{1}{l|}{0.729}           & 0.214           & 0.358           & 0.573           & \multicolumn{1}{l|}{0.632}           & 0.284                             & 0.403                             & 0.599                             & 0.651                             \\
\textbf{CubeRec}     & 0.478           & 0.494           & 0.654           & \multicolumn{1}{l|}{0.734}           & 0.422           & 0.440           & 0.602           & \multicolumn{1}{l|}{0.694}           & 0.073           & 0.101           & 0.174           & \multicolumn{1}{l|}{0.318}           & 0.078                             & 0.104                             & 0.188                             & 0.319                             \\

\textbf{S$^2$-HHGR}     & 0.527           & 0.562           & 0.785           & \multicolumn{1}{l|}{0.836}           & 0.484           & 0.515           & 0.761           & \multicolumn{1}{l|}{0.809}           & 0.374           & 0.392           & 0.628           & \multicolumn{1}{l|}{0.764}           & 0.434                             & 0.447                             & 0.688                             & 0.791                             \\

\textbf{SGGCF}     & 0.613           & 0.625           & 0.837           & \multicolumn{1}{l|}{0.860}           & 0.569           & 0.576           & 0.799           & \multicolumn{1}{l|}{0.832}           & 0.377           & 0.404           & 0.705           & \multicolumn{1}{l|}{0.814}           & 0.441                             & 0.462                             & 0.761                             & 0.843                             \\
\textbf{GroupIM}     & {\ul 0.642}     & {\ul 0.649}     & {\ul 0.854}     & \multicolumn{1}{l|}{{\ul 0.886}}     & {\ul 0.597}     & {\ul 0.606}     & {\ul 0.827}     & \multicolumn{1}{l|}{{\ul 0.874}}     & {\ul 0.388}     & {\ul 0.412}     & {\ul 0.711}     & \multicolumn{1}{l|}{{\ul 0.835}}     & {\ul 0.455}                       & {\ul 0.473}                       & {\ul 0.767}                       & {\ul 0.861}                       \\ \hline
\multicolumn{17}{c}{\textbf{Personality-Guided Aggregators}}                                                                                                                                                                                                                                                                                                                                                                                                 \\ \hline
\textbf{NCF+Big5}    & 0.433           & 0.468           & 0.521           & \multicolumn{1}{l|}{0.636}           & 0.395           & 0.429           & 0.587           & \multicolumn{1}{l|}{0.664}           & 0.275           & 0.293           & 0.423           & \multicolumn{1}{l|}{0.513}           & 0.324                             & 0.344                             & 0.486                             & 0.586                             \\
\textbf{PEGA}        & \textbf{0.670*} & \textbf{0.683*} & \textbf{0.874*} & \multicolumn{1}{l|}{\textbf{0.936*}} & \textbf{0.620*} & \textbf{0.634*} & \textbf{0.841*} & \multicolumn{1}{l|}{\textbf{0.914*}} & \textbf{0.421*} & \textbf{0.449*} & \textbf{0.721*} & \multicolumn{1}{l|}{\textbf{0.853*}} & \textbf{0.494*}                   & \textbf{0.514*}                   & \textbf{0.781*}                   & \textbf{0.884*}                   \\
\textit{VIP of PEGA} & 4.36\%          & 5.24\%          & 2.34\%          & \multicolumn{1}{l|}{5.64\%}          & 3.85\%          & 4.62\%          & 1.69\%          & \multicolumn{1}{l|}{4.58\%}          & 8.51\%          & 8.98\%          & 1.41\%          & \multicolumn{1}{l|}{2.16\%}          & 8.57\%                            & 8.67\%                            & 1.83\%                            & 2.67\%                            \\ \hline
\end{tabular}
\caption{Overall performance comparison on four datasets. (Note: * denotes the statistical significance for $p$-value \textless{} 0.05 compared to the best baseline, the boldface indicates the best model result of the dataset, and the underline indicates the second best model result of the dataset.)} 
\label{table3}
\end{table}
\end{landscape}
\restoregeometry

\begin{figure}[t]
  \centering
  \begin{subfigure}[b]{0.48\textwidth}
    \includegraphics[width=\textwidth]{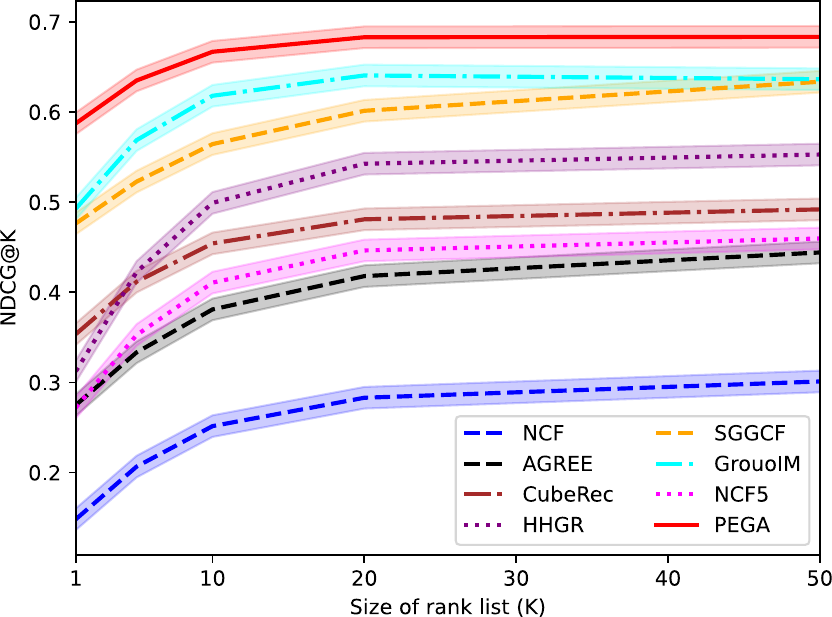}
    \caption{Amazon-Simi}
  \end{subfigure}
  \begin{subfigure}[b]{0.48\textwidth}
    \includegraphics[width=\textwidth]{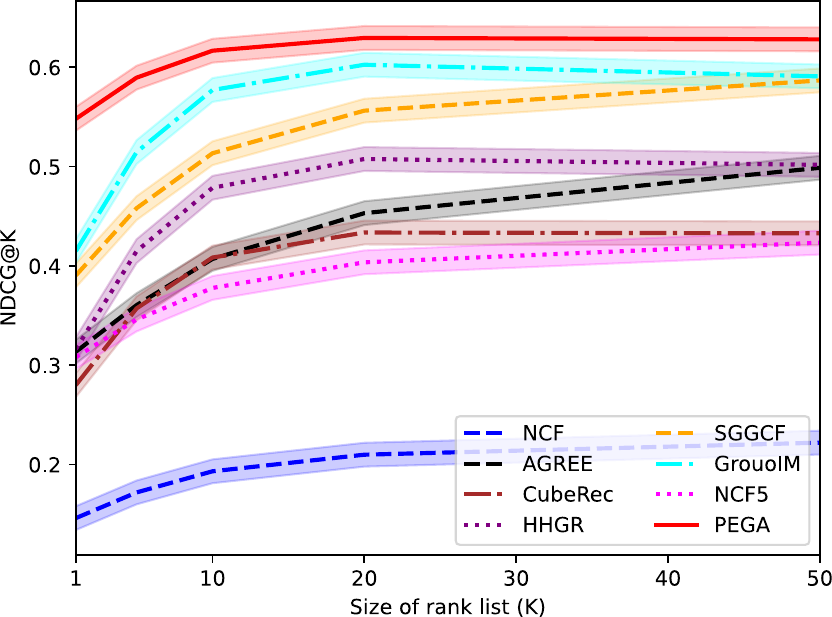}
    \caption{Amazon-Rand}
  \end{subfigure}
  \begin{subfigure}[b]{0.48\textwidth}
    \includegraphics[width=\textwidth]{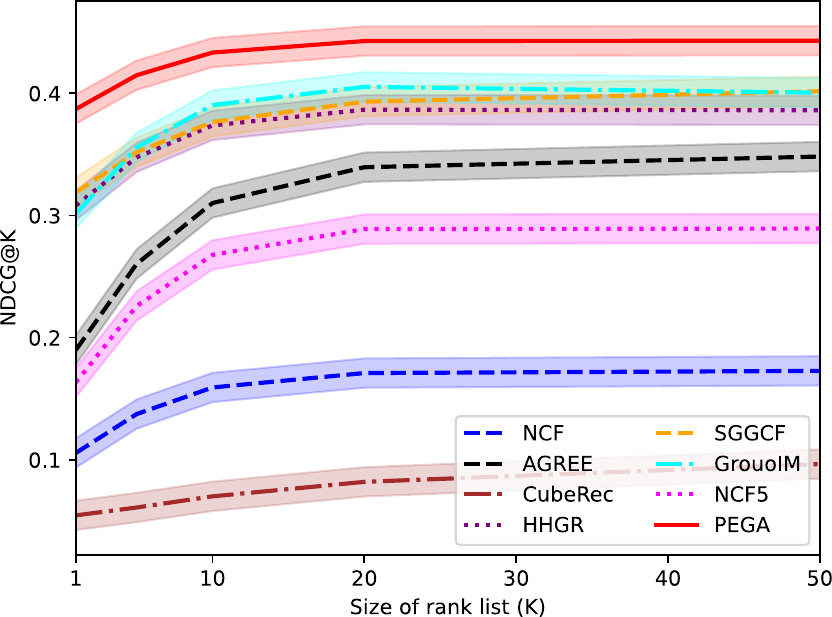}
    \caption{Yelp2017-2018}
  \end{subfigure}
  \begin{subfigure}[b]{0.48\textwidth}
    \includegraphics[width=\textwidth]{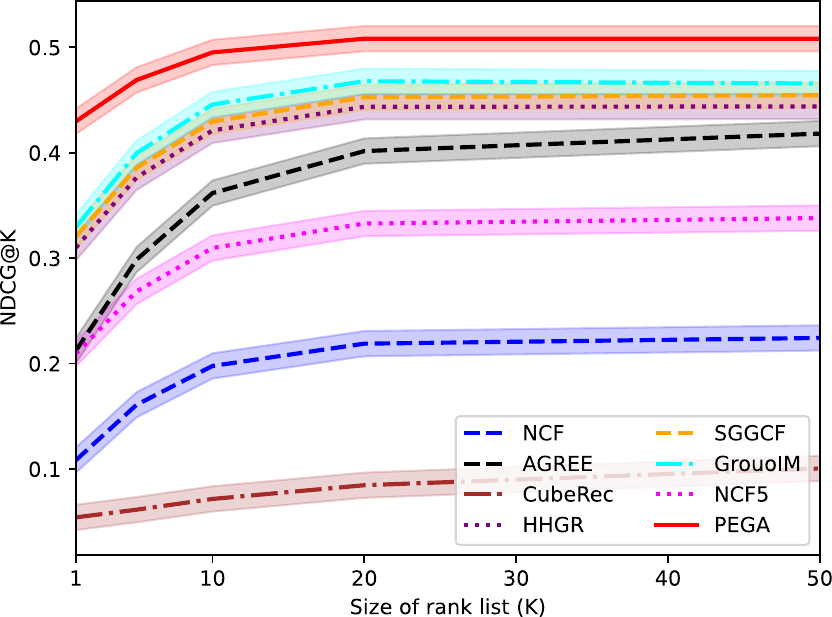}
    \caption{Yelp2019-2020}
  \end{subfigure}
  \caption{NDCG@K across size of rank list K. Variance bands indicate 95$\%$ confidence intervals over 10 random runs.}
  \label{figure5}
\end{figure}

\subsection{Result and Analysis}
\subsubsection{Overall Performance Comparison (RQ1)} \label{setcion5_4_1}
The overall comparison results of PEGA against state-of-the-art baselines are shown in \hyperref[table3]{Table 3}. It is observed that the proposed PEGA consistently outperforms all baselines on four datasets in terms of both R@K and N@K metrics. Compared to the strongest baseline GroupIM, PEGA shows a significant improvement ($p$-value \textless{} 0.05 via permutation test). 

Specifically, PEGA obtains the greatest advantage against predefined score aggregators on all datasets in terms of both metrics. For instance, when measured by R@20, the improvements against NCF-Best$\footnote{NCF-Best represents NCF-based methods with the best performance in different datasets, i.e., NCF+AVG in Yelp2017-2018 and Yelp2019-2020, NCF+LM in Amazon-Simi and Amazon-Rand.}$ are 77.6$\%$ in Amazon-Simi, 108.2$\%$ in Amazon-Rand, 169.0$\%$ in Yelp2017-2018 and 129.7$\%$ in Yelp2019-2020. It is possibly because PEGA adopts a data-driven strategy that dynamically models the group decision-making process and aggregates robust group preference representation rather than simply using hand-craft predefined strategies, which cannot adapt diversified scenarios in ephemeral groups.

By comparison, PEGA’s advantages against preference-guided aggregators are slightly smaller, but the improvements are still significant. Specifically, it can be seen in \hyperref[table3]{Table 3} that compared with the strongest preference-guided aggregator GroupIM, the proposed model PEGA achieves higher performance w.r.t. both N@K (e.g., VIP of N@20: 4.36$\%$ in Amazon-Simi vs. 3.85$\%$ in Amazon-Rand vs. 8.51$\%$ in Yelp2017-2018 vs. 8.57$\%$ in Yelp2019-2020) and R@K (e.g., VIP of R@20: 2.34$\%$ in Amazon-Simi vs. 1.69$\%$ in Amazon-Rand vs. 1.41$\%$ in Yelp2017-2018 vs. 1.83$\%$ in Yelp2019-2020). The reason is that PEGA simulates users’ importance not only by their history preferences but also by considering the influence of personality when they integrate into a new ephemeral group.

It is worth mentioning that PEGA significantly outperforms the NCF-Big5, another personality-guided aggregator, in terms of both N@K and R@K. For instance, in the Amazon-Simi dataset, the improvement of N@20 against NCF-Big5 is 54.7$\%$, while the average improvement on the other several datasets is 54.2$\%$. Similarly, the average improvement of R@20 against NCF-Big5 across all datasets is 59.5$\%$. This empirically validates that PEGA's proposed data-driven personality attention mechanism is more effective in realistically simulating the decision-making process of a new ephemeral group and providing more accurate recommendation results compared to a simple rule-based personality-guided aggregator. From another perspective, the superior performance of NCF-Big5 over NCF-Best (e.g., average N@20 improvement: 66.5$\%$) also demonstrates the superiority of guiding preference aggregation with personality in group recommendation tasks. 

In addition, regarding the performance of the model on different datasets, it is observed that all models achieve relatively lower abstract values of evaluation metrics on Yelp2017-2018 relative to Amazon-Simi, Amazon-Rand and Yelp2019-2020. As shown in \hyperref[table2]{Table 2}, the Yelp2017-2018 dataset owns the largest number of candidate items. Thus, more diverse individual preferences will lead to more complex group decision-making results, further exacerbating the group-level sparsity problem. However, PEGA still shows a significant improvement (e.g., average VIP of N@K: 8.75$\%$ and average VIP of R@K: 1.79$\%$). The possible reason is that, unlike preference-guided methods, PEGA aggregates diverse preferences with the guidance of consistent personality, which alleviates this problem and keeps stable superiority in such a challenging dataset. Another interesting observation from \hyperref[table3]{Table 3} is that all models perform better in Amazon-Simi than Amazon-Rand in both R@K and N@K. For example, the average values of R@20 and N@20 in Amazon-Simi are 0.640 and 0.442, while the values in Amazon-Rand are 0.612 and 0.403. It is reasonable since groups in Amazon-Simi consist of individuals with similar preferences, while groups in Amazon-Rand are structured randomly.

To further validate the effectiveness of the PEGA model, a detailed comparison of NDCG scores for all models across a range of K values (from 1 to 50) is performed. This choice is deliberate as NDCG focuses on giving higher ranks to positive instances in contrast to HR metrics, thus providing a fairer evaluation of the model's performance. \hyperref[figure5]{Fig. 5} shows the NDCG performance of all models over these varied K values. A thorough performance analysis reveals that PEGA consistently outperforms all baseline models across the range of K values. Notably, the advantage of PEGA becomes more apparent as K decreases. Compared to other models (e.g., GroupIM, SGGCF, and HHGR), PEGA exhibits a more gradual decline in NDCG values as K decreases, reflecting a smoother NDCG curve. This advantage stems from PEGA’s integration of personality traits into EGR, allowing for a more accurate simulation of group decision-making processes and better ranking of positive elements.

\begin{table}[t]
\small
\centering
\setlength{\tabcolsep}{1.8pt}
\begin{tabular}{lcc|cc|cc|cc}
\hline
\textbf{Dataset}   & \multicolumn{2}{c|}{\textbf{Amazon-Simi}} & \multicolumn{2}{c|}{\textbf{Amazon-Rand}} & \multicolumn{2}{c|}{\textbf{Yelp2017-2018}} & \multicolumn{2}{c}{\textbf{Yelp2019-2020}} \\
\textbf{Metric}    & N@20                & R@20                & N@20                & R@20                & N@20                 & R@20                 & N@20                 & R@20                \\ \hline
\textbf{BASE}      & 0.604               & 0.786               & 0.535               & 0.747               & 0.385                & 0.691                & 0.455                & 0.755               \\
\textbf{PEGA-nPER} & 0.630               & 0.828               & 0.580               & 0.783               & 0.402                & 0.708                & 0.472                & 0.767               \\
\textbf{PEGA-ATT}  & 0.642               & 0.834               & 0.588               & 0.805               & 0.407                & 0.711                & 0.477                & 0.769               \\
\textbf{PEGA-nPRE} & {\ul 0.653}         & {\ul 0.850}         & {\ul 0.609}         & {\ul 0.826}         & {\ul 0.417}          & {\ul 0.717}          & {\ul 0.491}          & {\ul 0.776}         \\ \hline
\textbf{PEGA}      & \textbf{0.670*}     & \textbf{0.874*}     & \textbf{0.620*}     & \textbf{0.841*}     & \textbf{0.421*}      & \textbf{0.721*}      & \textbf{0.494*}      & \textbf{0.781*}     \\ \hline
\end{tabular}
\caption{The comparison between PEGA and its variants. (Note: * denotes the statistical significance for $p$-value \textless{} 0.05 compared to the best variant, the boldface indicates the best model result of the dataset, and the underline indicates the second best model result of the dataset.)}
\label{table4}
\end{table}

\subsubsection{Ablation Study (RQ2)}
To verify the effectiveness of the main components of PEGA, ablation studies are conducted to evaluate several variants of PEGA. The results for all datasets, with N@20 and R@20 as metrics, are presented in \hyperref[table4]{Table 4}. 

Different components of PEGA mentioned in \hyperref[section4]{Section 4} are removed, and four variants are presented below: (1) \textbf{PEGA-nPER} sets the coefficient $\lambda$ in Eq. \hyperref[eq16]{(16)} as 1 to remove the personality attention mechanism; (i.e., no longer considering the influence of personality) (2) \textbf{PEGA-ATT} replaces the ``Group Personality'' represented by the average values of group members' personalities instead of hyper-rectangle and replaced the personality attention mechanism designed for hyper-rectangle with a simple attention mechanism. Specifically, the $\tilde{\alpha}(g, t)$ in Eq. \hyperref[eq7]{(7)} is replaced as
$\mathbf{H}^T \operatorname{Tanh}\left(\mathbf{W}_q AVG_g+\mathbf{W}_k P_t+\mathbf{b}\right)$, where $AVG_g$ is the average personality of group $g$'s members, while the other letters have the same meanings as in Eq. \hyperref[eq11]{(11)};
(3) \textbf{PEGA-nPRE} sets the coefficient $\lambda$ in Eq. \hyperref[eq16]{(16)} as 0 to remove the preference-based fine-tuning model (i.e., without considering users’ preference for candidate items); (4) \textbf{BASE} gives the same weight $\delta(g, t, i)=1$ in Eq. \hyperref[eq16]{(16)} to each group member $t$ in group $g$.

It can be seen from \hyperref[table4]{Table 4} that \textbf{PEGA} performs significantly better than the four variations across all datasets. Given the similarity in experimental results across the four datasets, key observations are illustrated using the Amazon-Simi dataset as an example. The notable observations are as follows: (1) PEGA obtains the greatest advantage
against \textbf{BASE} (e.g., VIP of N@20: 11.0$\%$ and VIP of R@20: 11.2$\%$). This is reasonable because \textbf{BASE} equally aggregates users’ preferences without any guidance; (2) PEGA performs slightly higher than \textbf{PEGA-nPRE} (e.g., VIP of N@20: 2.5$\%$ and VIP of R@20: 2.8$\%$) but obviously higher than \textbf{PEGA-nPER}(e.g., VIP of N@20: 6.4$\%$ and VIP of R@20: 5.6$\%$), implying that the weight of personality is more important than the weight of preference. Thus, it can be proved that personality would be more efficient in simulating users’ importance than preference, while preference would be more suitable to play a supporting role that fine-tunes the weight calculated by personality. (3) PEGA consistently outperforms \textbf{PEGA-ATT} across all datasets, highlighting that compared to a simple attention mechanism, PEGA's innovative design of ``Group Personality'' and personality attention mechanisms can more accurately simulate the importance of individual personalities within ``Group Personality''. In addition, \textbf{PEGA-ATT} demonstrates superiority over NCF+Big5 (e.g., VIP of N@20: 48.3$\%$ and VIP of R@20: 60.0$\%$), as shown in  \hyperref[table3]{Table 3}. This further exemplifies the superiority of PEGA's proposed data-driven personality attention mechanism over the rule-based personality-guided aggregation strategy.  

\begin{table}[t]
\small
\centering
\setlength{\tabcolsep}{2.3pt}
\begin{tabular}{lcc|cc|cc|cc}
\hline
\textbf{Dataset}    & \multicolumn{2}{l|}{\textbf{Amazon-Simi}} & \multicolumn{2}{l|}{\textbf{Amazon-Rand}} & \multicolumn{2}{l|}{\textbf{Yelp2017-2018}} & \multicolumn{2}{l}{\textbf{Yelp2019-2020}} \\
\textbf{Metric}     & \textbf{N@20}       & \textbf{R@20}       & \textbf{N@20}       & \textbf{R@20}       & \textbf{N@20}        & \textbf{R@20}        & \textbf{N@20}        & \textbf{R@20}       \\ \hline
\textbf{NCF-Best-P} & 0.296 $\uparrow$ & 0.508 $\uparrow$ & 0.212 $\uparrow$ & 0.419 $\uparrow$ & 0.179 $\uparrow$ & 0.287 $\uparrow$ & 0.235 $\uparrow$ & 0.363 $\uparrow$ \\
\textbf{AGREE-P}    & 0.285 $\downarrow$ & 0.607 $\downarrow$ & 0.377 $\downarrow$ & 0.656 $\downarrow$ & 0.210 $\downarrow$ & 0.568 $\downarrow$ & 0.276 $\downarrow$ & 0.590 $\downarrow$ \\
\textbf{CubeRec-P}  & 0.478 $\uparrow$ & 0.663 $\uparrow$ & 0.431 $\uparrow$ & 0.612 $\uparrow$ & 0.109 $\uparrow$ & 0.183 $\uparrow$ & 0.118 $\uparrow$ & 0.195 $\uparrow$ \\
\textbf{S$^2$-HHGR-P}  & 0.533 $\uparrow$ & 0.797 $\uparrow$ & 0.496 $\uparrow$ & 0.777 $\uparrow$ & 0.379 $\uparrow$ & 0.641 $\uparrow$ & 0.443 $\uparrow$ & 0.694 $\uparrow$ \\
\textbf{SGGCF-P}    & 0.629 $\uparrow$ & 0.847 $\uparrow$ & 0.585 $\uparrow$ & 0.816 $\uparrow$ & {\ul 0.383} $\uparrow$ & {\ul 0.710} $\uparrow$ & 0.449 $\uparrow$ & {\ul 0.768} $\uparrow$ \\
\textbf{GroupIM-P}   & {\ul 0.640} $\downarrow$ & {\ul 0.849} $\downarrow$ & {\ul 0.593} $\downarrow$ & {\ul 0.818} $\downarrow$ & 0.382 $\downarrow$ & 0.703 $\downarrow$ & {\ul 0.450} $\downarrow$ & 0.765 $\downarrow$ \\
\textbf{PEGA}       & \textbf{0.670*}     & \textbf{0.874*}     & \textbf{0.620*}     & \textbf{0.841*}     & \textbf{0.421*}      & \textbf{0.721*}      & \textbf{0.494*}      & \textbf{0.781*}     \\ \hline
\end{tabular}
\caption{The comparison between PEGA and baseline models with integrated personality information. (Note: * denotes the statistical significance for $p$-value \textless{} 0.05 compared to the best variant, the boldface indicates the best model result of the dataset, the underline indicates the second best model result of the dataset, and the upward and downward arrows respectively indicate the improvement or deterioration of the baseline model's performance compared to \hyperref[table3]{Table 3} after integrating personality information.)}
\label{table5}
\end{table}

\subsubsection{Effect of Personality Information (RQ3)}
Since PEGA employs implicit personality vectors derived from review data, for fairness, this auxiliary data ($P_u$ generated in \hyperref[section4_1]{Section 4.1}) is incorporated into all baseline models when learning user representations. The details of the changes to the baseline models are as follows: (1) \textbf{NCF-Best-P}: The best-performing NCF-based model on each dataset is selected, and the Big Five personality traits $P_u$ are concatenated with the user embeddings in the embedding layer of NCF; (2) \textbf{AGREE-P}: The Big-Five personality traits $P_u$ are integrated into the user embedding layer, and the dimensions of the user and item embeddings are aligned using a linear layer; (3) \textbf{CubeRec-P}: In the stage of pre-training user point embeddings, the Big Five personality traits $P_u$ are concatenated with the user embedding, and a linear layer is then used to align the dimensions of user and item embeddings; (4) \textbf{S$^2$-HHGR-P}: When constructing the user-level hypergraph, the Big Five personality traits $P_u$ are connected to the user embeddings, with a linear layer used to adjust the dimensions of user and item embeddings; (5) \textbf{SGGCF-P}: When initializing the user-centered graph, the Big-Five personality traits $P_u$ are integrated into the user node embedding; (6) \textbf{GroupIM-P}: During the pre-training phase of user embeddings, the Big Five personality traits $P_u$ are integrated into the user node embeddings through a linear layer.

From \hyperref[table5]{Table 5}, it can be observed that PEGA's performance on four datasets remains significantly higher than the baseline models that incorporate personality information derived from the review data (e.g., on the Amazon-Simi dataset, VIP of N@20: 4.7$\%$ and VIP of R@20: 2.9$\%$). This confirms the effectiveness and superiority of PEGA's overall model design, rather than relying solely on the additional use of user personality information. Simultaneously, among the baselines, the introduction of the Big Five personality traits $P_u$ improves the performance of over half of them (i.e., NCF-Best-P, CubeRec-P, S$^2$-HHGR-P, and SGGCF-P). This indicates that even a simple introduction of personality vectors at the user embedding level is beneficial for enhancing group recommendation systems' effectiveness. However, the performance of some models (i.e., AGREE-P and GroupIM-P) decreases after the introduction of personality vectors. This might be due to the fact that incorporating personality vectors at the user embedding level is not suitable for these frameworks, and it instead becomes noise. In general, experimental results demonstrate that personality is rather effective auxiliary information. If a suitable framework for incorporating personality can be designed (such as the design of ``Group Personality'' to address data sparsity issue and utilizing personality attention mechanism to simulate the importance of users in group decision-making), personality can play a more positive role.

\begin{figure}[t]
  \centering
  \begin{subfigure}{0.47\textwidth}
    \subcaptionbox*{$d$}{\includegraphics[width=\textwidth]{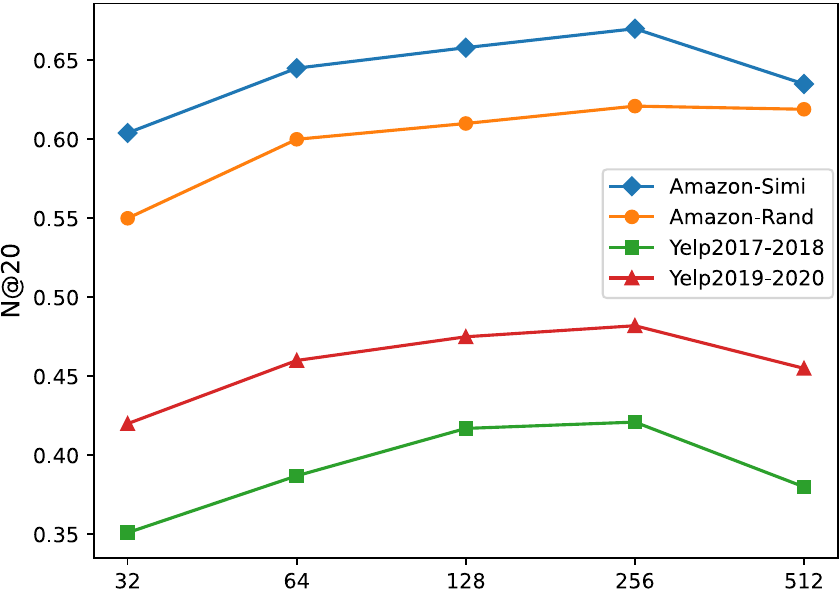}}
  \end{subfigure}
  \quad
  \begin{subfigure}{0.47\textwidth}
    \subcaptionbox*{$\gamma$}{\includegraphics[width=\textwidth]{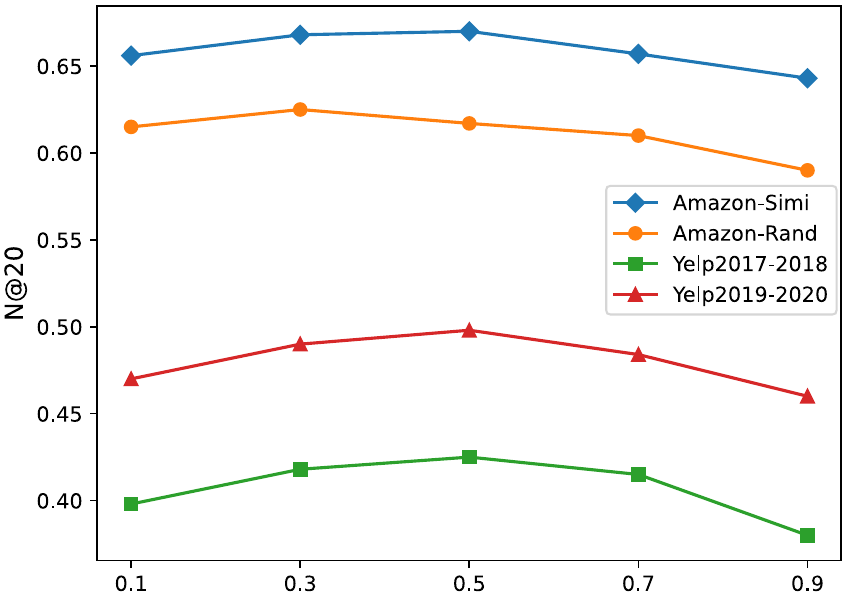}}
  \end{subfigure}
  \quad
  \begin{subfigure}{0.47\textwidth}
    \subcaptionbox*{$\mu$}{\includegraphics[width=\textwidth]{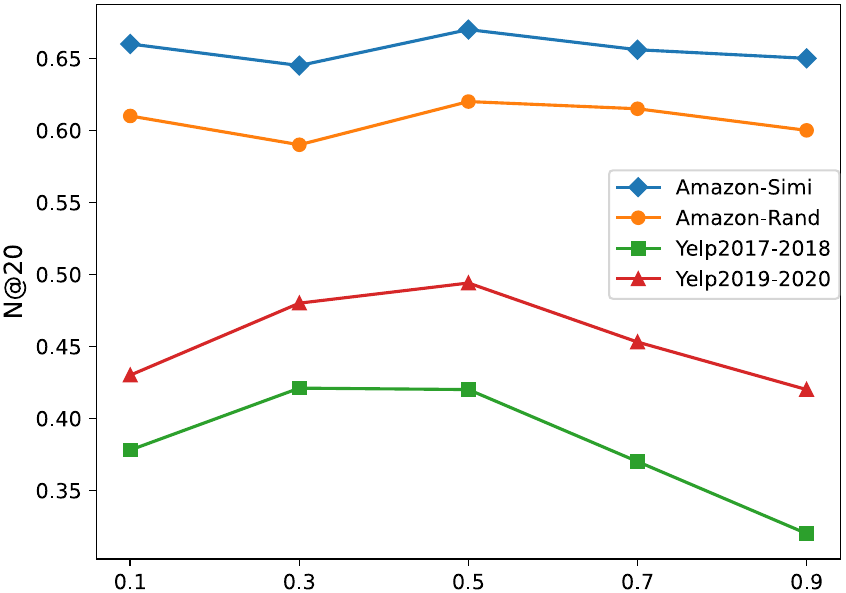}}
  \end{subfigure}
  \quad
  \begin{subfigure}{0.47\textwidth}
    \subcaptionbox*{$\lambda$}{\includegraphics[width=\textwidth]{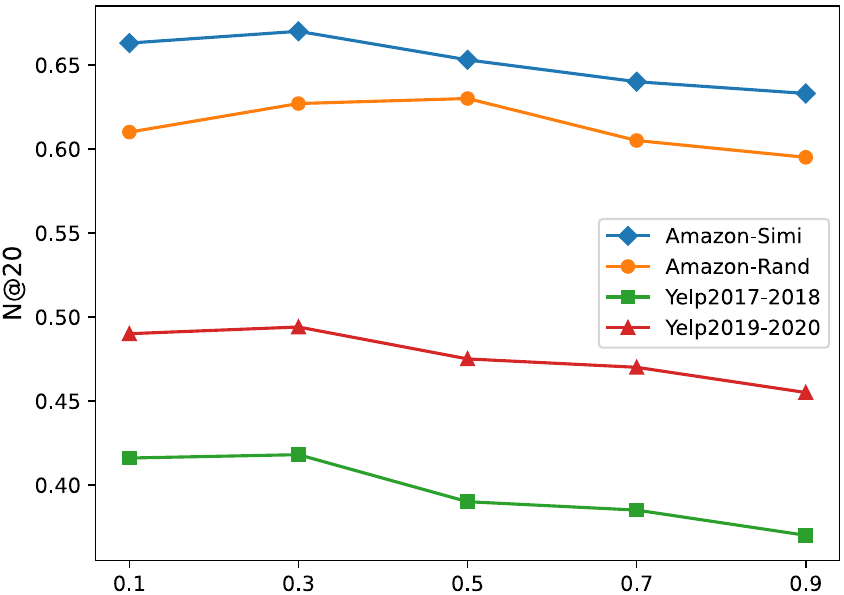}}
  \end{subfigure}
  \caption{The influence of different model hyperparameters.}
  \label{figure6}
\end{figure} 

\subsubsection{Hyperparameter Analysis (RQ4)}
RQ3 is answered by evaluating the latent dimension $d$, the balance coefficient of personality attention weights $\gamma$, the weight of the user ranking loss $\mu$, and the balance coefficient of personality and preference $\lambda$ on four datasets. Each hyperparameter is adjusted while others remain unchanged, and the results are reported in \hyperref[figure6]{Fig. 6}, with N@20 used for benchmarking. 

\textbf{Impact of $d$}. The latent dimension $d$ tuned in $\left\{32,64,128,256,512\right\}$. In general, PEGA benefits a larger latent dimension $d$ on all datasets. However, the performance decreases when $d$ is set too large (i.e., 512 in this case) due to overfitting.

\textbf{Impact of $\gamma$}. As mentioned in the previous works, the coefficient $\gamma$ in Eq. \hyperref[eq7]{(7)} plays a significant role in balancing personality attention weights. To analyze the impact of $\gamma$ on PEGA, the coefficient $\gamma$ is set among $\left\{0.1,0.3,0.5,0.7,0.9\right\}$. PEGA performance increases as $\gamma$ rises, until $\gamma$ reaches 0.5, after which it decreases with further increases in $\gamma$. The worst performance occurs when $\gamma=0.9$. This indicates that the relationship between individual personality and the main personality tone, as well as the relationship with extreme personalities, holds equal importance in PEGA. However, excessively high $\gamma$ values cause the model to overly prioritize the relationship between individual personality and the main personality tone, potentially leading to individuals losing their distinctiveness in personality, resulting in suboptimal outcomes.

\textbf{Impact of $\mu$}. The impact of $\mu \in \left\{0.1,0.3,0.5,0.7,0.9\right\} $, which controls the regularization effect of the user ranking loss, is also studied. The observations from \hyperref[figure6]{Fig. 6} indicate that the recommendation accuracy of PEGA initially increases as $\mu$ increases and then starts to decrease when $\mu$ exceeds 0.5. Lower $\mu$ values may result in underutilization of user-item interaction data and may not effectively alleviate the sparsity issue in group-item interaction data. On the other hand, setting $\mu$ too large may cause the model to prioritize individual recommendations and pay less attention to the group recommendation task.

\textbf{Impact of $\lambda$}. The parameter $\lambda$ is used for balancing the weight of the user’s personality and preference. When the value of $\lambda$ varies in $\left\{0.1,0.3,0.5,0.7,0.9\right\}$, the best performance can be observed when $\lambda=0.3$. In addition, the performance decreases as $\lambda$ increases in most scenarios, which testifies to the assumption that personality is more important than preference in the group decision-making process. It is noteworthy that personality and preference enjoy the same weight (i.e., $\lambda$=0.5) in Amazon-Rand. The possible reason is that groups in Amazon-Rand are too random (refer to the descriptions in \hyperref[section5_1]{Section 5.1}), which weakens the importance of the user’s personality. 

\begin{figure}[]
  \centering
  \begin{subfigure}[b]{0.48\textwidth}
    \includegraphics[width=\textwidth]{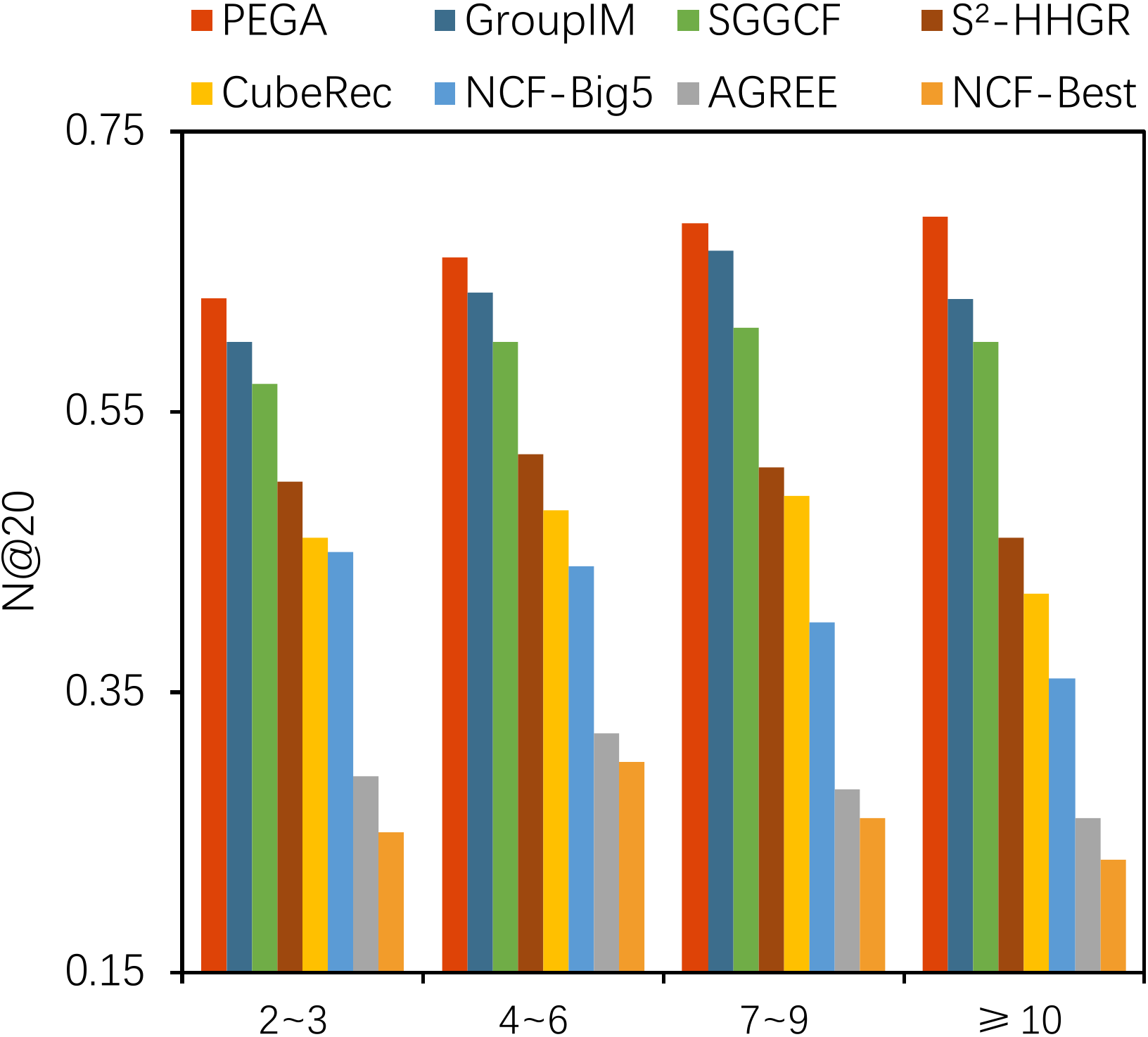}
    \caption{Amazon-Simi}
  \end{subfigure}
  \begin{subfigure}[b]{0.48\textwidth}
    \includegraphics[width=\textwidth]{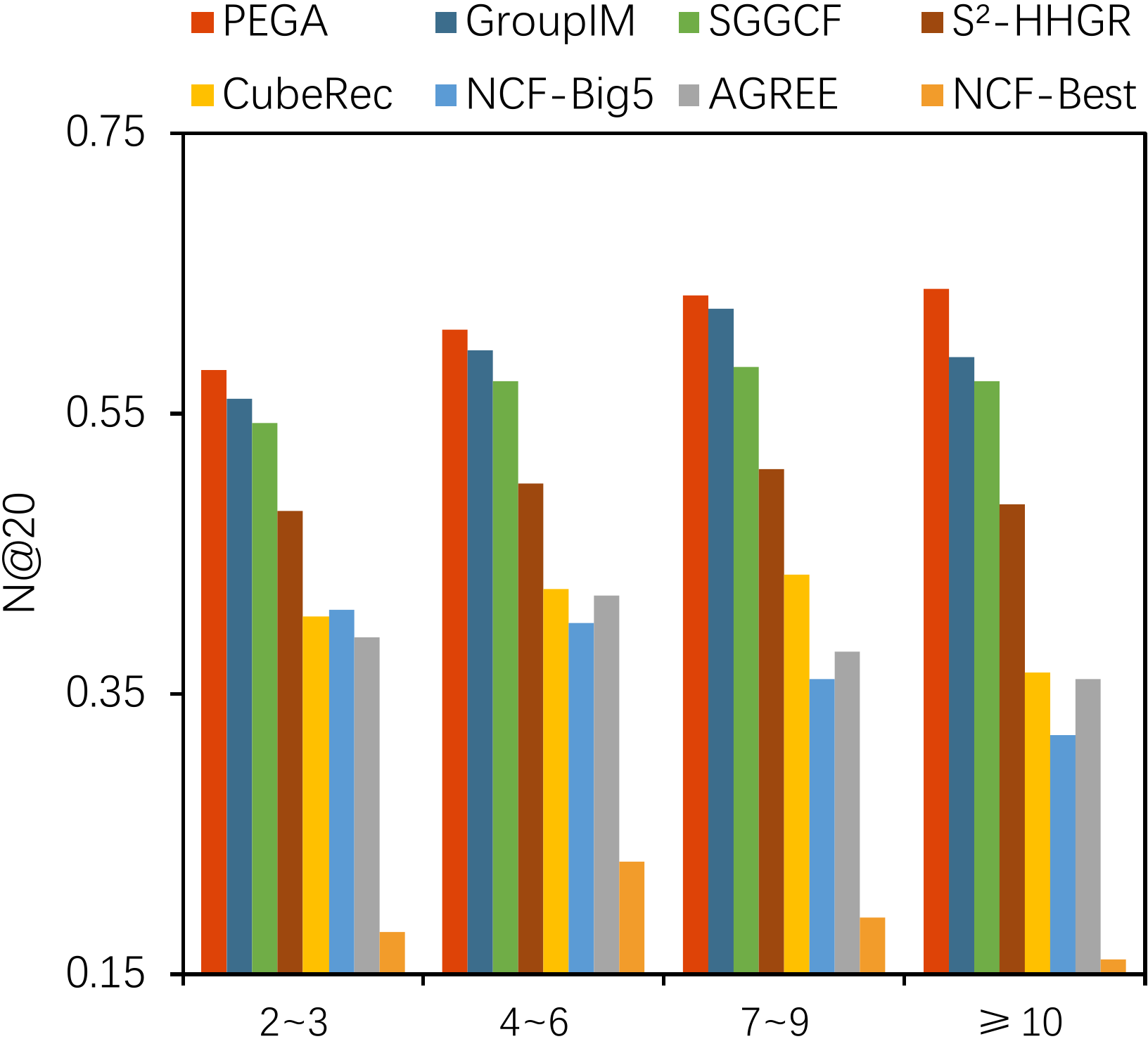}
    \caption{Amazon-Rand}
  \end{subfigure}
  \begin{subfigure}[b]{0.48\textwidth}
    \includegraphics[width=\textwidth]{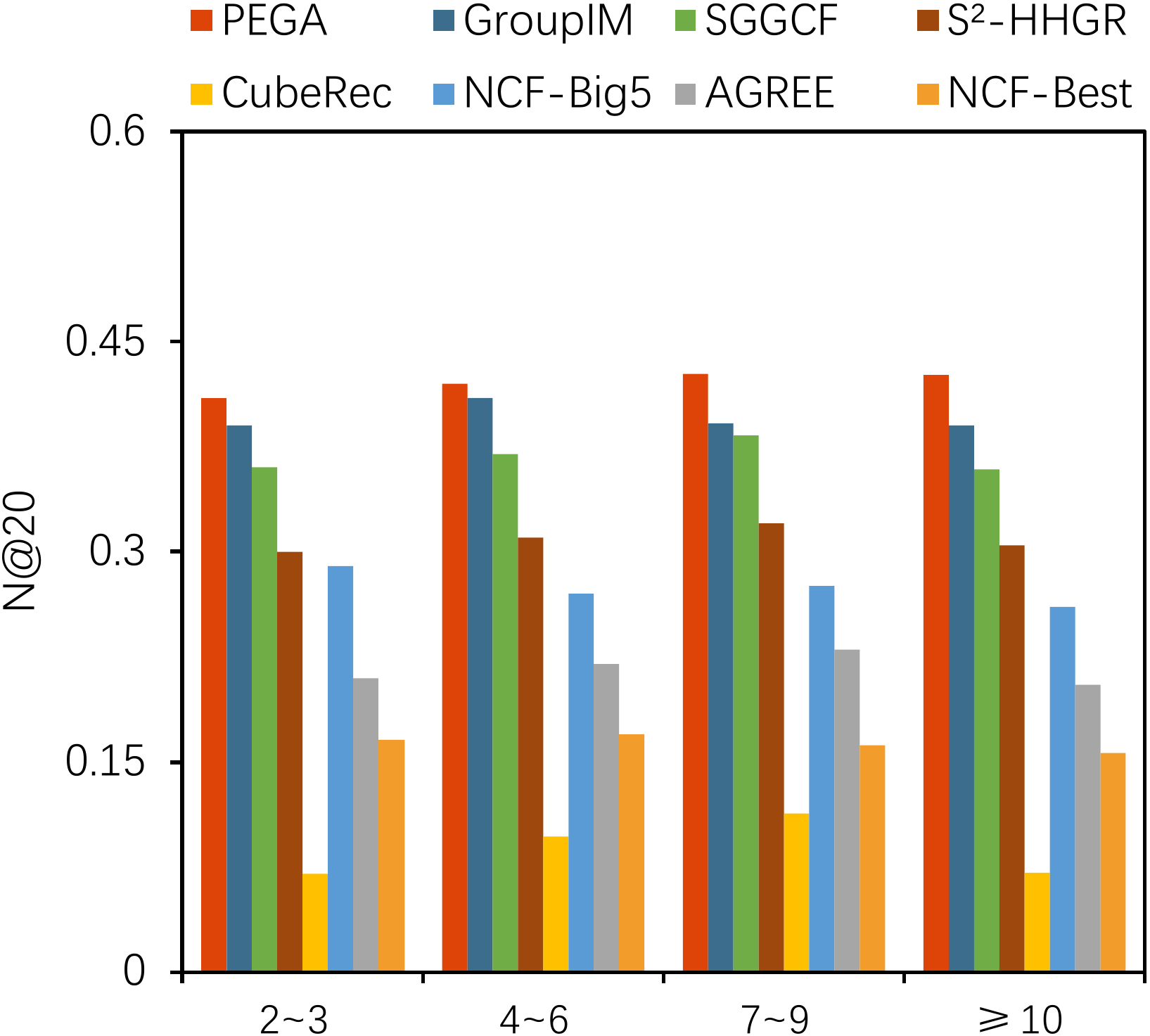}
    \caption{Yelp2017-2018}
  \end{subfigure}
  \begin{subfigure}[b]{0.48\textwidth}
    \includegraphics[width=\textwidth]{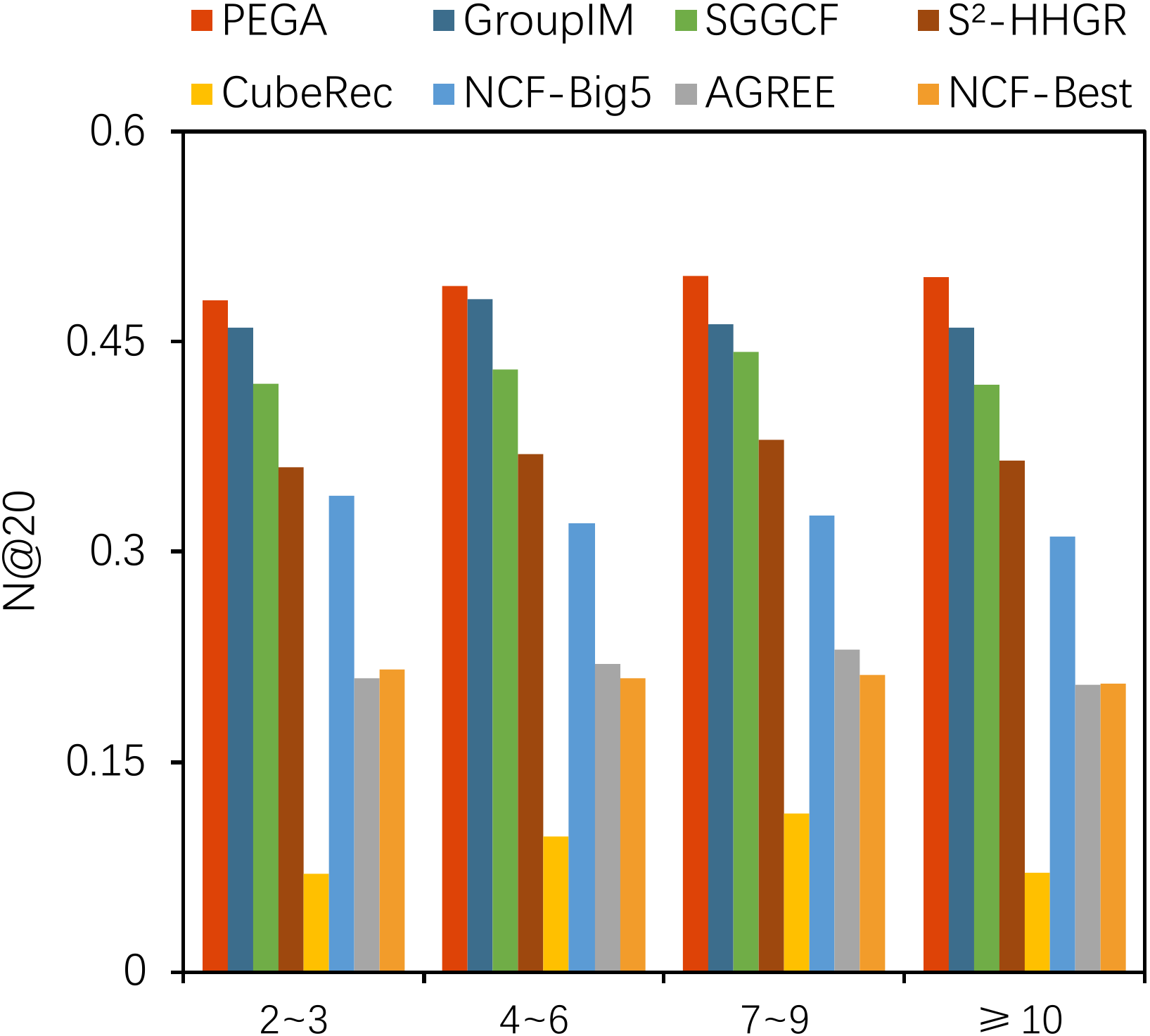}
    \caption{Yelp2019-2020}
  \end{subfigure}
  \caption{Performance across group size ranges.}
  \label{figure7}
\end{figure}

\subsubsection{Effect of Group Sizes (RQ5)}
To further investigate the impact of different group sizes on recommendation performance, group sizes are divided into four levels: 2-3, 4-6, 7-9, and 10 or more users, aligning with the approach in \citep{sankar2020groupim, chen2022thinking}. The recommendation results of N@20 scores across all datasets are depicted in \hyperref[figure7]{Fig. 7}. 

The results show that PEGA has strong robustness against different group sizes since it outperforms other models across the four datasets, which is consistent with the overall performance in \hyperref[table3]{Table 3}. An important observation is that the performances of most methods rise with the increase of the group size and the best performance mainly occurs when the group size is 4-9. It indicates that the group decision-making process in medium-sized groups is more regular and more learnable for group recommenders. On the contrary, the performance of NCF-Big5 decreases continuously as the group size increases, indicating that rule-based personality aggregation strategies are only applicable to small groups but cannot cope with complex ``Group Personality'' distributions. Meanwhile, most models suffer a noticeable performance decrease when the group size is too large (10 or more here), except PEGA. This is primarily associated with the preference noise caused by too diverse preferences of group members. However, PEGA studies the multiple distribution patterns of individual personalities in the group and assigns weights to users based on their personalities, which helps alleviate the impact of preference noise. This indicates that the proposed PEGA method has the potential to perform well across different group sizes.

\begin{figure}[t]
  \centering
  \includegraphics[width=0.8\linewidth]{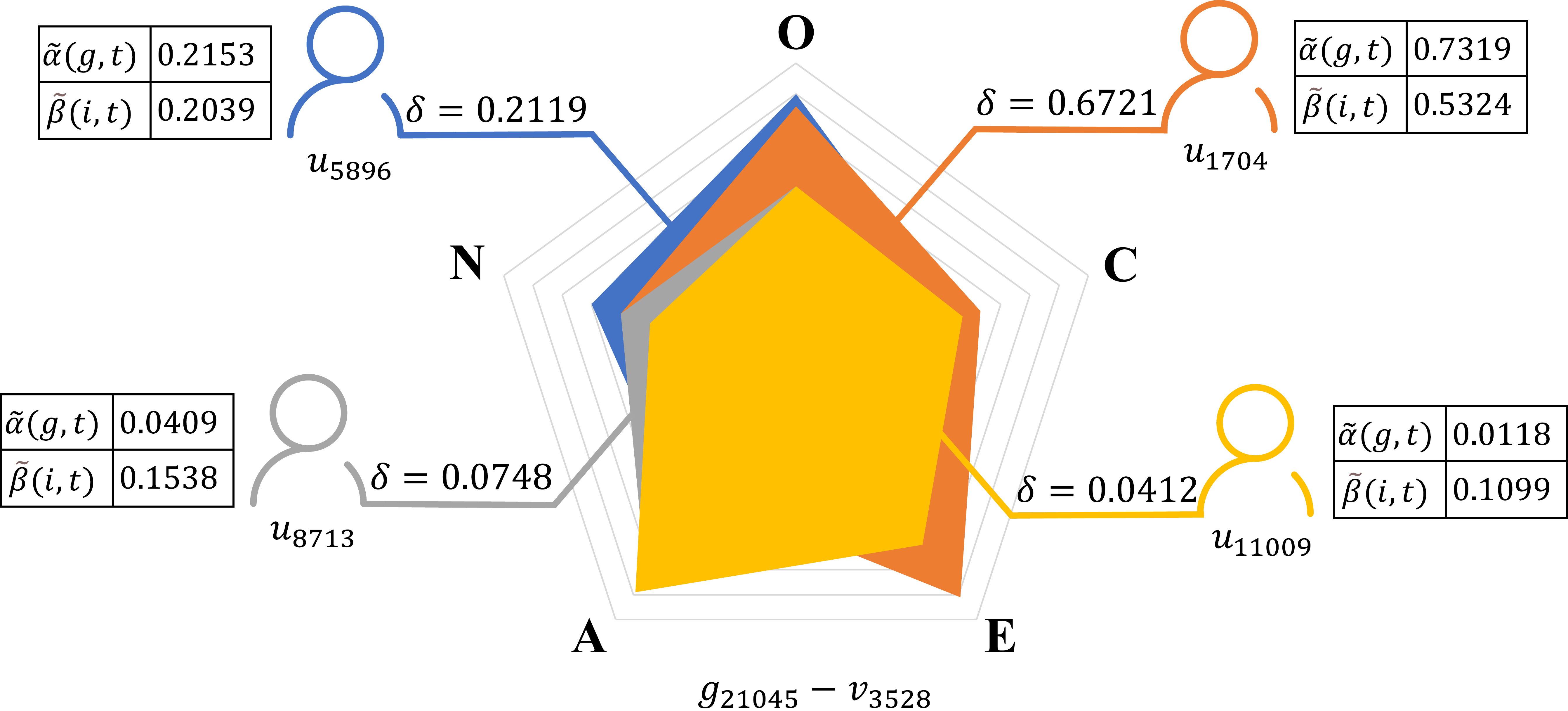}
  \caption{A visualized example from Yelp2017-2018, where $\delta$ denotes the weight of users' influence on the final group decision, $\tilde{\alpha}(g, t)$ stands for the weight of users' personality, and $\tilde{\beta}(i, t)$ signifies the weight of users' preferences.}
  \label{figure8}
\end{figure}

\subsubsection{Visualization of Individual Personality (RQ6)}
To intuitively demonstrate the impact of personality on group decision-making, the recommendation results for a randomly selected group, denoted as $g_{21045}$, are visualized using the Yelp2017-2018 dataset. In this particular case, the item $v_{3528}$ is correctly recommended to this group. For clarity, the implicit personality trait embeddings are mapped onto Big-Five radar charts. As shown in \hyperref[figure8]{Fig. 8}, the ``Group Personality'' of $g_{21045}$ can be described as high-level openness (O), high-level extroversion (E), and high-level Agreeableness (A). Under this personality contribution, the proposed model PEGA gives the user $u_{1704}$ the largest influence weight which is 0.6721, and sets $u_{5896}$ as second place with a weight of 0.2119 while other users’ influence weights are less than 0.1. Thus, the preference of $g_{21045}$ is primarily contributed by $u_{1704}$ and $u_{5896}$. Even though $u_{8713}$ and $u_{11009}$ do not express a preference for recommending item $v_{3528}$ (with their respective $\tilde{\beta}(i, t)$ values being 0.15 and 0.11), group $g_{21045}$ still opts for $v_{3528}$ as its final choice. Additionally, it's worth noting that both $u_{1704}$ and $u_{5896}$ exhibit higher scores in \textit{Openness}, \textit{Extraversion}, and \textit{Neuroticism}, indicating their outgoing and confident nature. As a result, they would be more willing to express their preference. Conversely, other users demonstrate relatively higher levels of \textit{Agreeableness}, suggesting their easygoing nature and inclination to follow others' suggestions. This example explicitly illustrates that in a group comprising assertive and easygoing individuals, the assertive ones tend to exert a greater influence on the group's decision-making process. 

\begin{figure}[t]
  \centering
  \begin{subfigure}[b]{0.48\textwidth}
    \includegraphics[width=\textwidth]{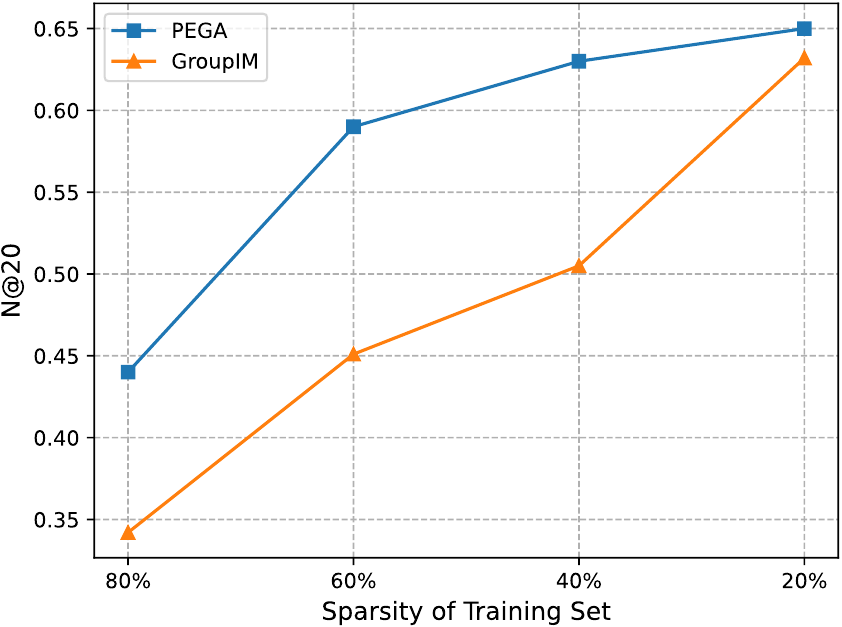}
    \caption{Amazon-Simi}
  \end{subfigure}
  \quad
  \begin{subfigure}[b]{0.48\textwidth}
    \includegraphics[width=\textwidth]{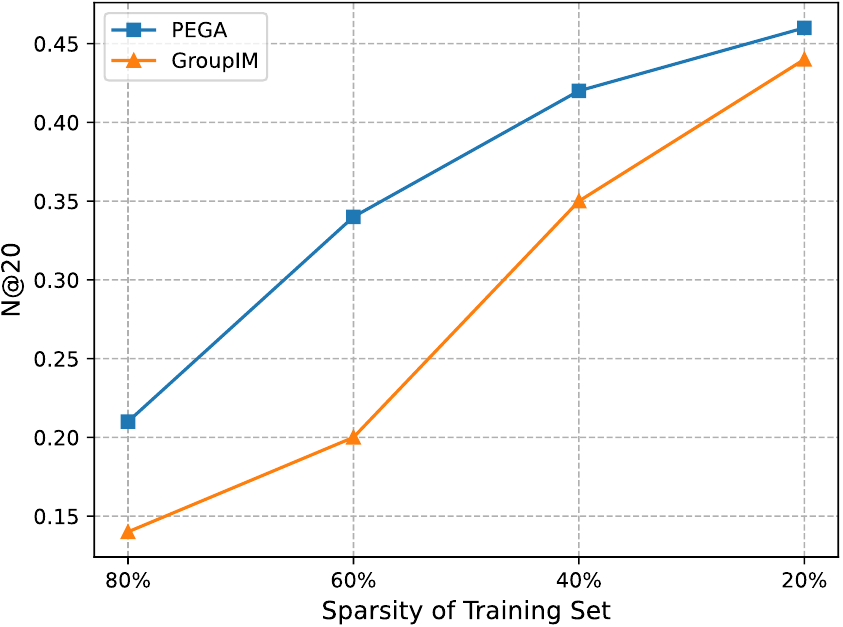}
    \caption{Yelp2019-2020}
  \end{subfigure}
  \caption{Performance across different data sparsity levels.}
  \label{figure9}
\end{figure}

\subsubsection{Visualization of ``Group Personality'' (RQ7)}\label{setcion5_4_7}
To investigate the potential of ``Group Personality'' in addressing the challenge of ephemeral group data sparsity, performance tests are conducted on PEGA and the strongest baseline, GroupIM, across different levels of data sparsity. Specifically, $\left\{80\%, 60\%, 40\%, 20\%\right\}$ of the training set is randomly removed, corresponding to the sparsity levels on the x-axis of \hyperref[figure9]{Fig. 9}, $\left\{80\%, 60\%, 40\%, 20\%\right\}$ respectively. The recommendation results for N@20 scores on the Amazon-Simi and Yelp2019-2020 datasets are presented. Notably, PEGA outperforms GroupIM, especially in scenarios with severe data sparsity. For instance, using only 40$\%$ of the training data from the Yelp2019-2020 dataset, PEGA achieves 68.8$\%$ of the best performance in N@20, whereas GroupIM only reaches 43.9$\%$ of the best performance.

\begin{figure}[t]
  \centering
  \begin{subfigure}[b]{0.32\textwidth}
    \includegraphics[width=\textwidth]{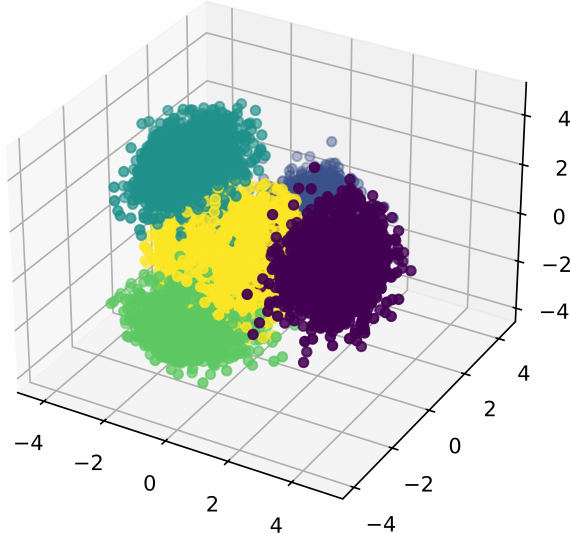}
    \caption{``Group Personality'' of PEGA}
    \label{figure10a}
  \end{subfigure}
  \begin{subfigure}[b]{0.32\textwidth}
    \includegraphics[width=\textwidth]{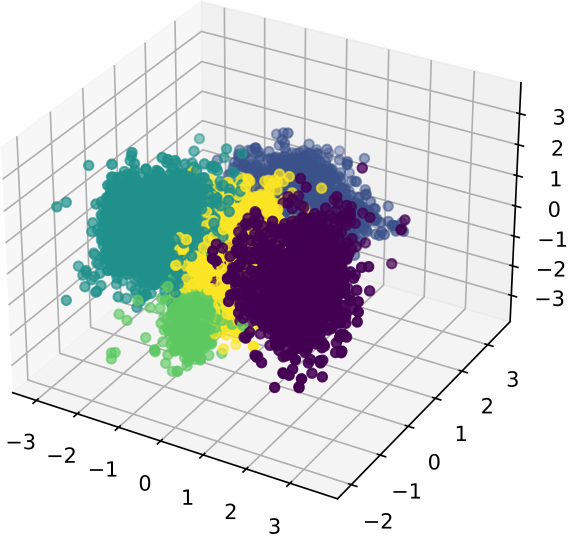}
    \caption{Group Embedding of PEGA}
    \label{figure10b}
  \end{subfigure}
  \begin{subfigure}[b]{0.32\textwidth}
    \includegraphics[width=\textwidth]{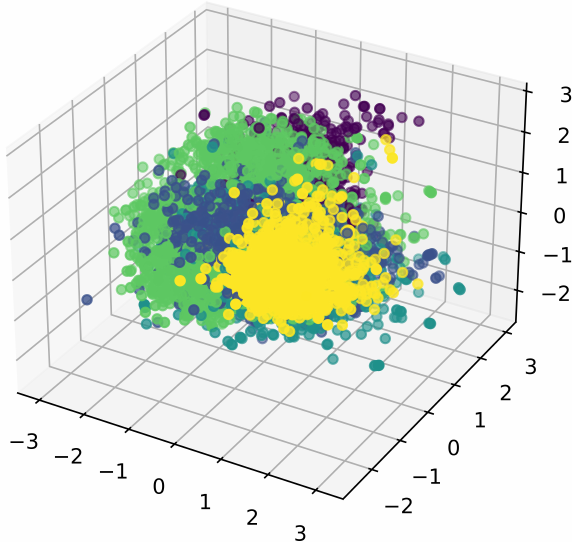}
    \caption{Group Embedding of GroupIM}
    \label{figure10c}
  \end{subfigure}
  \caption{A visualized example of learned embeddings by PEGA and GroumIM.}
  \label{figure10}
\end{figure}

Furthermore, the embeddings learned by PEGA and GroupIM at a sparsity level of 40$\%$ are visualized in \hyperref[figure10]{Fig. 10}. For clarity, only the first three dimensions of the embeddings are utilized after conducting principal component analysis (PCA), and the results are visualized by clustering. The visualization indicates that, even with just around half of the training data, PEGA acquires a certain understanding of the distribution of group personalities (\hyperref[figure10a]{Fig. 10(a)}). Leveraging this knowledge, the personality-guided aggregation of the group preference embedding (\hyperref[figure10b]{Fig. 10(b)}) demonstrates faster and more effective convergence compared to GroupIM, which neglects personality factors (\hyperref[figure10c]{Fig. 10(c)}). This example clearly demonstrates that utilizing ``Group Personality'' to guide group preference aggregation helps alleviate the issue of sparse data in ephemeral groups.

\section{Discussion}\label{section6}
\subsection{Summary of Experimental Findings}
In summary, the overall performance comparison results (\hyperref[table3]{Table 3}) demonstrate that the proposed personality-guided preference aggregator PEGA outperforms existing preference-based methods \citep{cao2018attentive, chen2022thinking, zhang2021double, Li2023SelfSupervisedGG, sankar2020groupim} in terms of top-k recommendation metrics. A detailed performance comparison using the NDCG (1-50) metric (\hyperref[figure5]{Fig. 5}) further validates the superiority and effectiveness of PEGA. Additionally, experiments demonstrate the role of user personality in EGR scenarios. In particular, ablation experiments (\hyperref[table4]{Table 4}) show that personality factors are more effective than preference factors in simulating user importance, while preferences are better suited to play a supplementary role in fine-tuning the weights computed from personality. On the other hand, the proposed ``Group Personality'' concept proves to be more valuable in addressing group-level data sparsity issues (\hyperref[figure9]{Fig. 9}), as it quickly understands the distribution of group personalities from limited data, which helps to more effectively guide the convergence of preference embeddings (\hyperref[figure10]{Fig. 10}).

Beyond these overall results, this research brings several practical implications for EGR. First of all, the proposed review-based personality extraction module can be directly integrated into existing baseline models, improving their performance in EGR scenarios (\hyperref[table5]{Table 5}). This demonstrates the broad applicability of personality information for enhancing EGR task performance, consistent with prior research \citep{abolghasemi2022personality, Rossi2016SocialUA}. In contrast, whereas methods that obtain user personalities through questionnaires are limited, the approach of extracting implicit user personalities from comments can be flexibly incorporated into any group recommendation model. As a result, this advancement significantly enhances the role of personality in optimizing EGR systems. 

Secondly, the case study reveals that individuals with varying personalities perform differently in group decision-making processes. Specifically, individuals with higher scores in the Big-Five personality dimensions of \textit{Openness}, \textit{Extraversion}, and \textit{Neuroticism} are more likely to express their preferences, while those with higher \textit{Agreeableness} scores tend to follow the suggestions of others in the group (\hyperref[figure8]{Fig. 8}). This indicates that more confident individuals often exert a greater influence on the group's decision-making process. Nguyen et al. \citep{nguyen2019conflict} also find similar results. They use the Thomas-Kilmann Conflict Model to simulate users' conflict resolution styles and discover that individuals adopting the avoiding conflict style experience the greatest personal loss (deviation from their own preferences).

Thirdly, impact analysis demonstrates that recommendation performance in group recommendation systems varies with group size (\hyperref[figure7]{Fig. 7}). Most baseline models achieve optimal performance when the group size is between 4 and 9 but experience a noticeable performance decline when the group size exceeds 10. This finding is consistent with the work of Sankar et al. \citep{sankar2020groupim} and Chen et al. \citep{chen2022thinking}. However, unlike these studies, PEGA still maintains its performance for large groups, as it examines the multiple distribution patterns of individual personalities within the group and assigns weights based on user personality. This approach helps reduce preference noise and ensures robustness across different group sizes.  

\subsection{Limitations}\label{section6_2}  
From the perspective of the dataset, although personality traits have been shown to enhance EGR performance, the limited types of data features in the chosen dataset mean that other influencing factors—such as the price of recommended items, geographic location, weather conditions, and timing of group activities—are not considered. Future work could incorporate these additional features and use sensitivity analysis techniques \citep{asheghi2020updating, yeung2010sensitivity, zhang2019novel} to select optimal features, thereby improving the effectiveness of EGR. 

Furthermore, the dataset used in this study is static, as noted in previous EGR work \citep{sankar2020groupim, Deng2021KnowledgeAwareGR, zhou2023multi}. This static nature may result in the model's inability to maintain optimal performance due to the lack of continuous updates. Thus, designing an automated data collection and processing pipeline to create a dynamically updated database \citep{abbaszadeh2024normalizing} would be highly valuable in future research.

From the perspective of model structure design, the current method benefits from its review-based personality extraction module. However, this approach may be limited when users have sparse personal comment records. An alternative could be to extract implicit information from other sources, such as the number of user friends and social media profiles \citep{lima2014multi, tandera2017personality}. Additionally, a mixed-method approach, incorporating both qualitative (e.g., questionnaires) and quantitative (e.g., social media profiles) personality data \citep{ekal2022defi}, might be beneficial.

In terms of model optimization, the joint training strategy depicted in Eq. \hyperref[eq19]{(19)} trains both $\mathcal{L}{user}$ and $\mathcal{L}{group}$ together, which surpasses the performance of existing models. However, research \citep{hosseini2022prediction, abbaszadeh2021landslide, zou20183d} suggests that a divide-and-conquer task decomposition strategy could enhance overall task performance. Incorporating a block-based model design into EGR models and improving user and group representations through two-stage training \citep{huang2021novel} may potentially improve recommendation performance.

\section{Conclusions}
Existing preference-guided group recommenders lack consideration of other influencing factors in the real group decision-making process and suffer from the group-level data sparsity problem. In this study, a novel personality-guided preference aggregator (PEGA) for EGR is proposed. This method implicitly captures users' personality traits from their review texts and uses a personality attention mechanism to guide group preference aggregation. To address data sparsity in ephemeral groups, the concept of ``Group Personality'' is defined using a hyper-rectangle, representing individual integration into the group in a generalizable way.  

The accuracy of PEGA's top-k recommendations is evaluated using Recall. In the Yelp2019-2020 dataset, PEGA shows a 129.7$\%$ improvement in R@20 over the best predefined score aggregator (NCF-Best) and a 1.83$\%$ improvement over the best preference-guided aggregator (GroupIM). Additionally, PEGA's ranking quality is assessed using NDCG, with N@20 improving by 130.8$\%$ compared to NCF-Best and by 8.57$\%$ compared to GroupIM, reflecting PEGA's superior recommendation performance. Moreover, a detailed discussion on the impact of personality on the performance of EGR models is presented. (1) More than half of the baseline models (7 out of 9) show improved performance with the integration of the implicit personality feature. On the Yelp2017-2018 dataset, the average improvement in N@20 is 6.3$\%$, highlighting the importance and versatility of personality features in group recommenders. (2) In more sparse scenarios, using only 40$\%$ of the training data from the Yelp2019-2020 dataset, PEGA achieves 68.8$\%$ of the best performance in N@20, while GroupIM reaches only 43.9$\%$ of the best performance, demonstrating the effectiveness of the designed ``Group Personality'' structure in alleviating the issue of sparse group data. 

The results of this study on the importance and effectiveness of personality in EGR can be effectively integrated into existing models and future designs of EGR systems. This integration enhances the ability to more accurately and realistically simulate the group decision-making process.

\bibliographystyle{elsarticle-num}  
\bibliography{reference}

\begin{thebibliography}{10}
\expandafter\ifx\csname url\endcsname\relax
  \def\url#1{\texttt{#1}}\fi
\expandafter\ifx\csname urlprefix\endcsname\relax\def\urlprefix{URL }\fi
\expandafter\ifx\csname href\endcsname\relax
  \def\href#1#2{#2} \def\path#1{#1}\fi

\bibitem{quintarelli2019efficiently}
E.~Quintarelli, E.~Rabosio, L.~Tanca, Efficiently using contextual influence to recommend new items to ephemeral groups, Information Systems 84 (2019) 197--213.
\newblock \href {https://doi.org/https://doi.org/10.1016/j.is.2019.05.003} {\path{doi:https://doi.org/10.1016/j.is.2019.05.003}}.

\bibitem{ceh2022performance}
E.~Ceh-Varela, H.~Cao, H.~W. Lauw, Performance evaluation of aggregation-based group recommender systems for ephemeral groups, ACM Transactions on Intelligent Systems and Technology (TIST) 13~(6) (2022) 1--26.
\newblock \href {https://doi.org/https://doi.org/10.1145/3542804} {\path{doi:https://doi.org/10.1145/3542804}}.

\bibitem{ghazarian2015enhancing}
S.~Ghazarian, M.~A. Nematbakhsh, Enhancing memory-based collaborative filtering for group recommender systems, Expert systems with applications 42~(7) (2015) 3801--3812.
\newblock \href {https://doi.org/https://doi.org/10.1016/j.eswa.2014.11.042} {\path{doi:https://doi.org/10.1016/j.eswa.2014.11.042}}.

\bibitem{cao2019social}
D.~Cao, X.~He, L.~Miao, G.~Xiao, H.~Chen, J.~Xu, Social-enhanced attentive group recommendation, IEEE Transactions on Knowledge and Data Engineering (TKDE) 33~(3) (2019) 1195--1209.
\newblock \href {https://doi.org/https://doi.org/10.1109/TKDE.2019.2936475} {\path{doi:https://doi.org/10.1109/TKDE.2019.2936475}}.

\bibitem{huang2020efficient}
Z.~Huang, X.~Xu, H.~Zhu, M.~Zhou, An efficient group recommendation model with multiattention-based neural networks, IEEE Transactions on Neural Networks and Learning Systems 31~(11) (2020) 4461--4474.
\newblock \href {https://doi.org/https://doi.org/10.1109/TNNLS.2019.2955567} {\path{doi:https://doi.org/10.1109/TNNLS.2019.2955567}}.

\bibitem{wang2021socially}
P.~Wang, L.~Li, R.~Wang, G.~Xu, J.~Zhang, Socially-driven multi-interaction attentive group representation learning for group recommendation, Pattern Recognition Letters 145 (2021) 74--80.
\newblock \href {https://doi.org/https://doi.org/10.1016/j.patrec.2021.02.007} {\path{doi:https://doi.org/10.1016/j.patrec.2021.02.007}}.

\bibitem{yu2023collaborative}
L.~Yu, Y.~Leng, D.~Zhang, S.~He, Collaborative group embedding and decision aggregation based on attentive influence of individual members: A group recommendation perspective, Decision Support Systems 165 (2023) 113894.
\newblock \href {https://doi.org/https://doi.org/10.1016/j.dss.2022.113894} {\path{doi:https://doi.org/10.1016/j.dss.2022.113894}}.

\bibitem{he2022h3rec}
Z.~He, C.-Y. Chow, J.-D. Zhang, K.-Y. Lam, H3rec: Higher-order heterogeneous and homogeneous interaction modeling for group recommendations of web services, IEEE Transactions on Services Computing (TSC) 16~(2) (2022) 1212--1224.
\newblock \href {https://doi.org/https://doi.org/10.1109/TSC.2022.3180163} {\path{doi:https://doi.org/10.1109/TSC.2022.3180163}}.

\bibitem{jiang2023ktpgn}
X.~Jiang, H.~Sun, Y.~Chen, L.~He, Ktpgn: Novel event-based group recommendation method considering implicit social trust and knowledge propagation, Information Sciences 642 (2023) 119159.
\newblock \href {https://doi.org/https://doi.org/10.1016/j.ins.2023.119159} {\path{doi:https://doi.org/10.1016/j.ins.2023.119159}}.

\bibitem{abolghasemi2024graph}
R.~Abolghasemi, E.~H. Viedma, P.~Engelstad, Y.~Djenouri, A.~Yazidi, A graph neural approach for group recommendation system based on pairwise preferences, Information Fusion 107 (2024) 102343.
\newblock \href {https://doi.org/https://doi.org/10.1016/j.inffus.2024.102343} {\path{doi:https://doi.org/10.1016/j.inffus.2024.102343}}.

\bibitem{sankar2020groupim}
A.~Sankar, Y.~Wu, Y.~Wu, W.~Zhang, H.~Yang, H.~Sundaram, Groupim: A mutual information maximization framework for neural group recommendation, in: Proceedings of the International ACM SIGIR Conference on Research and Development in Information Retrieval (SIGIR), 2020, pp. 1279--1288.
\newblock \href {https://doi.org/https://doi.org/10.1145/3397271.3401116} {\path{doi:https://doi.org/10.1145/3397271.3401116}}.

\bibitem{Li2023SelfSupervisedGG}
K.~Li, C.-D. Wang, J.-H. Lai, H.~Yuan, Self-supervised group graph collaborative filtering for group recommendation, in: Proceedings of the ACM International Conference on Web Search and Data Mining (WSDM), 2023, pp. 69--77.
\newblock \href {https://doi.org/https://doi.org/10.1145/3539597.3570400} {\path{doi:https://doi.org/10.1145/3539597.3570400}}.

\bibitem{chen2022thinking}
T.~Chen, H.~Yin, J.~Long, Q.~V.~H. Nguyen, Y.~Wang, M.~Wang, Thinking inside the box: Learning hypercube representations for group recommendation, in: Proceedings of the International ACM SIGIR Conference on Research and Development in Information Retrieval (SIGIR), 2022, pp. 1664--1673.
\newblock \href {https://doi.org/https://doi.org/10.1145/3477495.3532066} {\path{doi:https://doi.org/10.1145/3477495.3532066}}.

\bibitem{abolghasemi2022personality}
R.~Abolghasemi, P.~Engelstad, E.~Herrera-Viedma, A.~Yazidi, A personality-aware group recommendation system based on pairwise preferences, Information Sciences 595 (2022) 1--17.
\newblock \href {https://doi.org/https://doi.org/10.1016/j.ins.2022.02.033} {\path{doi:https://doi.org/10.1016/j.ins.2022.02.033}}.

\bibitem{santos2011personality}
R.~Santos, G.~Marreiros, C.~Ramos, J.~Neves, J.~Bulas-Cruz, Personality, emotion, and mood in agent-based group decision making, IEEE Intelligent Systems 26~(06) (2011) 58--66.
\newblock \href {https://doi.org/https://doi.org/10.1109/MIS.2011.92} {\path{doi:https://doi.org/10.1109/MIS.2011.92}}.

\bibitem{wu2018personalizing}
W.~Wu, L.~Chen, Y.~Zhao, Personalizing recommendation diversity based on user personality, User Modeling and User-Adapted Interaction (UMUAI) 28~(3) (2018) 237--276.
\newblock \href {https://doi.org/https://doi.org/10.1007/s11257-018-9205-x} {\path{doi:https://doi.org/10.1007/s11257-018-9205-x}}.

\bibitem{dhelim2020personality}
S.~Dhelim, H.~Ning, N.~Aung, R.~Huang, J.~Ma, Personality-aware product recommendation system based on user interests mining and metapath discovery, IEEE Transactions on Computational Social Systems 8~(1) (2020) 86--98.
\newblock \href {https://doi.org/https://doi.org/10.1109/TCSS.2020.3037040} {\path{doi:https://doi.org/10.1109/TCSS.2020.3037040}}.

\bibitem{mairesse2007using}
F.~Mairesse, M.~A. Walker, M.~R. Mehl, R.~K. Moore, Using linguistic cues for the automatic recognition of personality in conversation and text, Journal of Artificial Intelligence Research (JAIR) 30 (2007) 457--500.
\newblock \href {https://doi.org/https://doi.org/10.1613/jair.2349} {\path{doi:https://doi.org/10.1613/jair.2349}}.

\bibitem{shin2009socially}
C.~Shin, W.~Woo, Socially aware tv program recommender for multiple viewers, IEEE Transactions on Consumer Electronics 55~(2) (2009) 927--932.
\newblock \href {https://doi.org/10.1109/TCE.2009.5174476} {\path{doi:10.1109/TCE.2009.5174476}}.

\bibitem{boratto2011state}
L.~Boratto, S.~Carta, State-of-the-art in group recommendation and new approaches for automatic identification of groups, Information Retrieval and Mining in Distributed Environments (2011) 1--20\href {https://doi.org/https://doi.org/10.1007/978-3-642-16089-9_1} {\path{doi:https://doi.org/10.1007/978-3-642-16089-9_1}}.

\bibitem{liu2012exploring}
X.~Liu, Y.~Tian, M.~Ye, W.-C. Lee, Exploring personal impact for group recommendation, in: Proceedings of the ACM International Conference on Information and Knowledge Management (CIKM), 2012, pp. 674--683.
\newblock \href {https://doi.org/https://doi.org/10.1145/2396761.2396848} {\path{doi:https://doi.org/10.1145/2396761.2396848}}.

\bibitem{yuan2014generative}
Q.~Yuan, G.~Cong, C.-Y. Lin, Com: a generative model for group recommendation, in: Proceedings of the ACM SIGKDD International Conference on Knowledge Discovery and Data Mining (SIGKDD), 2014, pp. 163--172.
\newblock \href {https://doi.org/https://doi.org/10.1145/2623330.2623616} {\path{doi:https://doi.org/10.1145/2623330.2623616}}.

\bibitem{cao2018attentive}
D.~Cao, X.~He, L.~Miao, Y.~An, C.~Yang, R.~Hong, Attentive group recommendation, in: Proceedings of the International ACM SIGIR Conference on Research and Development in Information Retrieval (SIGIR), 2018, pp. 645--654.
\newblock \href {https://doi.org/https://doi.org/10.1145/3209978.3209998} {\path{doi:https://doi.org/10.1145/3209978.3209998}}.

\bibitem{yin2019social}
H.~Yin, Q.~Wang, K.~Zheng, Z.~Li, J.~Yang, X.~Zhou, Social influence-based group representation learning for group recommendation, in: International Conference on Data Engineering (ICDE), 2019, pp. 566--577.
\newblock \href {https://doi.org/https://doi.org/10.1109/ICDE.2019.00057} {\path{doi:https://doi.org/10.1109/ICDE.2019.00057}}.

\bibitem{Deng2021KnowledgeAwareGR}
Z.~Deng, C.~Li, S.~Liu, W.~Ali, J.~Shao, Knowledge-aware group representation learning for group recommendation, in: International Conference on Data Engineering (ICDE), IEEE, 2021, pp. 1571--1582.
\newblock \href {https://doi.org/https://doi.org/10.1109/ICDE51399.2021.00139} {\path{doi:https://doi.org/10.1109/ICDE51399.2021.00139}}.

\bibitem{zhang2021double}
J.~Zhang, M.~Gao, J.~Yu, L.~Guo, J.~Li, H.~Yin, Double-scale self-supervised hypergraph learning for group recommendation, in: Proceedings of the ACM International Conference on Information and Knowledge Management (CIKM), 2021, pp. 2557--2567.
\newblock \href {https://doi.org/https://doi.org/10.1145/3459637.3482426} {\path{doi:https://doi.org/10.1145/3459637.3482426}}.

\bibitem{wang2021cross}
H.~Wang, Y.~Zuo, H.~Li, J.~Wu, Cross-domain recommendation with user personality, Knowledge-Based Systems 213 (2021) 106664.
\newblock \href {https://doi.org/https://doi.org/10.1016/j.knosys.2020.106664} {\path{doi:https://doi.org/10.1016/j.knosys.2020.106664}}.

\bibitem{quijano2017make}
L.~Quijano-Sanchez, C.~Sauer, J.~A. Recio-Garcia, B.~Diaz-Agudo, Make it personal: a social explanation system applied to group recommendations, Expert Systems with Applications 76 (2017) 36--48.
\newblock \href {https://doi.org/https://doi.org/10.1016/j.eswa.2017.01.045} {\path{doi:https://doi.org/10.1016/j.eswa.2017.01.045}}.

\bibitem{zheng2018exploring}
Y.~Zheng, Exploring user roles in group recommendations: A learning approach, in: Adjunct Publication of the Conference on User Modeling, Adaptation and Personalization, 2018, pp. 49--52.
\newblock \href {https://doi.org/https://doi.org/10.1145/3213586.3226192} {\path{doi:https://doi.org/10.1145/3213586.3226192}}.

\bibitem{alves2023group}
P.~Alves, H.~Martins, P.~Saraiva, J.~Carneiro, P.~Novais, G.~Marreiros, Group recommender systems for tourism: how does personality predict preferences for attractions, travel motivations, preferences and concerns?, User Modeling and User-Adapted Interaction 33~(5) (2023) 1141--1210.
\newblock \href {https://doi.org/https://doi.org/10.1007/s11257-023-09361-2} {\path{doi:https://doi.org/10.1007/s11257-023-09361-2}}.

\bibitem{recio2009personality}
J.~A. Recio-Garcia, G.~Jimenez-Diaz, A.~A. Sanchez-Ruiz, B.~Diaz-Agudo, Personality aware recommendations to groups, in: ACM Recommender Systems Conference (Recsys), 2009, pp. 325--328.
\newblock \href {https://doi.org/https://doi.org/10.1145/1639714.1639779} {\path{doi:https://doi.org/10.1145/1639714.1639779}}.

\bibitem{kilmann1977developing}
R.~H. Kilmann, K.~W. Thomas, Developing a forced-choice measure of conflict-handling behavior: The ``mode" instrument, Educational and Psychological Measurement 37~(2) (1977) 309--325.
\newblock \href {https://doi.org/https://doi.org/10.1177/001316447703700204} {\path{doi:https://doi.org/10.1177/001316447703700204}}.

\bibitem{mccrae1992introduction}
R.~R. McCrae, O.~P. John, An introduction to the five-factor model and its applications, Journal of Personality 60~(2) (1992) 175--215.
\newblock \href {https://doi.org/https://doi.org/10.1111/j.1467-6494.1992.tb00970.x} {\path{doi:https://doi.org/10.1111/j.1467-6494.1992.tb00970.x}}.

\bibitem{pennebaker1999linguistic}
J.~W. Pennebaker, L.~A. King, Linguistic styles: language use as an individual difference., Journal of personality and social psychology 77~(6) (1999) 1296.
\newblock \href {https://doi.org/https://doi.org/10.1037/0022-3514.77.6.1296} {\path{doi:https://doi.org/10.1037/0022-3514.77.6.1296}}.

\bibitem{kwantes2016assessing}
P.~J. Kwantes, N.~Derbentseva, Q.~Lam, O.~Vartanian, H.~H. Marmurek, Assessing the big five personality traits with latent semantic analysis, Personality and Individual Differences 102 (2016) 229--233.
\newblock \href {https://doi.org/https://doi.org/10.1016/j.paid.2016.07.010} {\path{doi:https://doi.org/10.1016/j.paid.2016.07.010}}.

\bibitem{wang2020leverage}
X.~Wang, H.~Zhang, L.~Cao, L.~Feng, Leverage social media for personalized stress detection, in: Proceedings of the ACM International Conference on Multimedia (MM), 2020, pp. 2710--2718.
\newblock \href {https://doi.org/https://doi.org/10.1145/3394171.3413596} {\path{doi:https://doi.org/10.1145/3394171.3413596}}.

\bibitem{aizawa2003information}
A.~Aizawa, An information-theoretic perspective of tf--idf measures, Information Processing \& Management 39~(1) (2003) 45--65.
\newblock \href {https://doi.org/https://doi.org/10.1016/S0306-4573(02)00021-3} {\path{doi:https://doi.org/10.1016/S0306-4573(02)00021-3}}.

\bibitem{he2020lightgcn}
X.~He, K.~Deng, X.~Wang, Y.~Li, Y.~Zhang, M.~Wang, Lightgcn: Simplifying and powering graph convolution network for recommendation, in: Proceedings of the International ACM SIGIR Conference on Research and Development in Information Retrieval (SIGIR), 2020, pp. 639--648.
\newblock \href {https://doi.org/https://doi.org/10.1145/3397271.3401063} {\path{doi:https://doi.org/10.1145/3397271.3401063}}.

\bibitem{rendle2009bpr}
S.~Rendle, C.~Freudenthaler, Z.~Gantner, L.~Schmidt-Thieme, Bpr: Bayesian personalized ranking from implicit feedback, in: Proceedings of the Conference on Uncertainty in Artificial Intelligence (UAI), 2009, pp. 452--461.
\newblock \href {https://doi.org/https://dl.acm.org/doi/10.5555/1795114.1795167} {\path{doi:https://dl.acm.org/doi/10.5555/1795114.1795167}}.

\bibitem{zhou2023multi}
W.~Zhou, Z.~Huang, C.~Wang, Y.~Chen, A multi-graph neural group recommendation model with meta-learning and multi-teacher distillation, Knowledge-Based Systems (2023) 110731\href {https://doi.org/https://doi.org/10.1016/j.knosys.2023.110731} {\path{doi:https://doi.org/10.1016/j.knosys.2023.110731}}.

\bibitem{pearson1896vii}
K.~Pearson, Vii. mathematical contributions to the theory of evolution.—iii. regression, heredity, and panmixia, Philosophical Transactions of the Royal Society of London. Series A, containing papers of a mathematical or physical character~(187) (1896) 253--318.
\newblock \href {https://doi.org/https://doi.org/10.1098/rsta.1896.0007} {\path{doi:https://doi.org/10.1098/rsta.1896.0007}}.

\bibitem{guo2021hierarchical}
L.~Guo, H.~Yin, T.~Chen, X.~Zhang, K.~Zheng, Hierarchical hyperedge embedding-based representation learning for group recommendation, ACM Transactions on Information Systems (TOIS) 40~(1) (2021) 1--27.
\newblock \href {https://doi.org/https://doi.org/10.1145/3457949} {\path{doi:https://doi.org/10.1145/3457949}}.

\bibitem{he2017neural}
X.~He, L.~Liao, H.~Zhang, L.~Nie, X.~Hu, T.-S. Chua, Neural collaborative filtering, in: Proceedings of the International Conference on World Wide Web (WWW), 2017, pp. 173--182.
\newblock \href {https://doi.org/https://doi.org/10.1145/3038912.3052569} {\path{doi:https://doi.org/10.1145/3038912.3052569}}.

\bibitem{Rossi2016SocialUA}
S.~Rossi, F.~Cervone, Social utilities and personality traits for group recommendation: A pilot user study., in: Proceedings of the International Conference on Agents and Artificial Intelligence (ICAART), 2016, pp. 38--46.
\newblock \href {https://doi.org/https://doi.org/10.5220/0005709600380046} {\path{doi:https://doi.org/10.5220/0005709600380046}}.

\bibitem{nichols2002nonparametric}
T.~E. Nichols, A.~P. Holmes, Nonparametric permutation tests for functional neuroimaging: a primer with examples, Human Brain Mapping 15~(1) (2002) 1--25.
\newblock \href {https://doi.org/https://doi.org/10.1002/hbm.1058} {\path{doi:https://doi.org/10.1002/hbm.1058}}.

\bibitem{nguyen2019conflict}
T.~N. Nguyen, F.~Ricci, A.~Delic, D.~Bridge, Conflict resolution in group decision making: insights from a simulation study, User Modeling and User-Adapted Interaction (UMUAI) 29 (2019) 895--941.
\newblock \href {https://doi.org/https://doi.org/10.1007/s11257-019-09240-9} {\path{doi:https://doi.org/10.1007/s11257-019-09240-9}}.

\bibitem{asheghi2020updating}
R.~Asheghi, S.~A. Hosseini, M.~Saneie, A.~A. Shahri, Updating the neural network sediment load models using different sensitivity analysis methods: a regional application, Journal of Hydroinformatics 22~(3) (2020) 562--577.
\newblock \href {https://doi.org/https://doi.org/10.2166/hydro.2020.098} {\path{doi:https://doi.org/10.2166/hydro.2020.098}}.

\bibitem{yeung2010sensitivity}
D.~S. Yeung, I.~Cloete, D.~Shi, W.~wY~Ng, Sensitivity analysis for neural networks, Springer, 2010.
\newblock \href {https://doi.org/https://doi.org/10.1007/978-3-642-02532-7} {\path{doi:https://doi.org/10.1007/978-3-642-02532-7}}.

\bibitem{zhang2019novel}
P.~Zhang, A novel feature selection method based on global sensitivity analysis with application in machine learning-based prediction model, Applied Soft Computing 85 (2019) 105859.
\newblock \href {https://doi.org/https://doi.org/10.1016/j.asoc.2019.105859} {\path{doi:https://doi.org/10.1016/j.asoc.2019.105859}}.

\bibitem{abbaszadeh2024normalizing}
A.~Abbaszadeh~Shahri, C.~Shan, S.~Larsson, F.~Johansson, Normalizing large scale sensor-based mwd data: An automated method toward a unified database, Sensors 24~(4) (2024) 1209.
\newblock \href {https://doi.org/https://doi.org/10.3390/s24041209} {\path{doi:https://doi.org/10.3390/s24041209}}.

\bibitem{lima2014multi}
A.~C.~E. Lima, L.~N. De~Castro, A multi-label, semi-supervised classification approach applied to personality prediction in social media, Neural Networks 58 (2014) 122--130.
\newblock \href {https://doi.org/https://doi.org/10.1016/j.neunet.2014.05.020} {\path{doi:https://doi.org/10.1016/j.neunet.2014.05.020}}.

\bibitem{tandera2017personality}
T.~Tandera, D.~Suhartono, R.~Wongso, Y.~L. Prasetio, et~al., Personality prediction system from facebook users, Procedia computer science 116 (2017) 604--611.
\newblock \href {https://doi.org/https://doi.org/10.1016/j.procs.2017.10.016} {\path{doi:https://doi.org/10.1016/j.procs.2017.10.016}}.

\bibitem{ekal2022defi}
H.~H. Ekal, S.~N. Abdul-wahab, Defi governance and decision-making on blockchain, Mesopotamian Journal of Computer Science 2022 (2022) 10--17.
\newblock \href {https://doi.org/https://doi.org/10.58496/MJCSC/2022/003} {\path{doi:https://doi.org/10.58496/MJCSC/2022/003}}.

\bibitem{hosseini2022prediction}
S.~A. Hosseini, A.~Abbaszadeh~Shahri, R.~Asheghi, Prediction of bedload transport rate using a block combined network structure, Hydrological Sciences Journal 67~(1) (2022) 117--128.
\newblock \href {https://doi.org/https://doi.org/10.1080/02626667.2021.2003367} {\path{doi:https://doi.org/10.1080/02626667.2021.2003367}}.

\bibitem{abbaszadeh2021landslide}
A.~Abbaszadeh~Shahri, F.~Maghsoudi~Moud, Landslide susceptibility mapping using hybridized block modular intelligence model, Bulletin of engineering geology and the environment 80 (2021) 267--284.
\newblock \href {https://doi.org/https://doi.org/10.1007/s10064-020-01922-8} {\path{doi:https://doi.org/10.1007/s10064-020-01922-8}}.

\bibitem{zou20183d}
B.-J. Zou, Y.-D. Guo, Q.~He, P.-B. Ouyang, K.~Liu, Z.-L. Chen, 3d filtering by block matching and convolutional neural network for image denoising, Journal of Computer Science and Technology 33 (2018) 838--848.
\newblock \href {https://doi.org/https://doi.org/10.1007/s11390-018-1859-7} {\path{doi:https://doi.org/10.1007/s11390-018-1859-7}}.

\bibitem{huang2021novel}
Z.~Huang, Y.~Liu, C.~Zhan, C.~Lin, W.~Cai, Y.~Chen, A novel group recommendation model with two-stage deep learning, IEEE Transactions on Systems, Man, and Cybernetics: Systems 52~(9) (2021) 5853--5864.
\newblock \href {https://doi.org/https://doi.org/10.1109/TSMC.2021.3131349} {\path{doi:https://doi.org/10.1109/TSMC.2021.3131349}}.

\end{thebibliography}

\end{document}